\documentclass[a4paper,fleqn,usenatbib]{mnras}
\usepackage{lscape}
\usepackage{hyperref}

\usepackage[T1]{fontenc}
\usepackage{ae,aecompl}

\usepackage{graphicx}	
\usepackage{amsmath}	
\usepackage{amssymb}	

\title[Primeval VLMS and brown dwarfs I]{Primeval very low-mass stars and brown dwarfs -- I.  Six new L subdwarfs, classification and atmospheric properties}

\author[Z. H. Zhang et al.]{Z. H. Zhang,$^{1,2}$\thanks{E-mail:
zenghuazhang@gmail.com}\thanks{Based on observations made with ESO telescopes at the La Silla Paranal Observatory under programmes 088.C-0048, 091.C-0452, 094.C-0202, 096.C-0130.} 
D. J. Pinfield,$^{3}$  M. C. G\'{a}lvez-Ortiz,$^{4}$  B. Burningham,$^{3}$   
\newauthor 
 N. Lodieu,$^{1,2}$ F. Marocco,$^{3}$  A. J. Burgasser,$^{5}$  A. C. Day-Jones,$^{3}$  F. Allard,$^{6}$  
\newauthor 
H. R. A. Jones,$^{3}$  D. Homeier,$^{7}$     J. Gomes,$^{3}$ R. L. Smart$^{8}$ 
\\
$^{1}$Instituto de Astrof{\'i}sica de Canarias, E-38205 La Laguna, Tenerife, Spain \\
$^{2}$Deptment Astrof{\'i}sica, Universidad de La Laguna, E-38206 La Laguna, Tenerife, Spain \\
$^{3}$Centre for Astrophysics Research, Science and Technology Research Institute, University of Hertfordshire, Hatfield AL10 9AB, UK \\
$^{4}$Centro de Astrobiolog{\'i}a (CSIC-INTA), Ctra. Ajalvir km 4, E-28850 Torrej{\'o}n de Ardoz, Madrid, Spain \\
$^{5}$Center for Astrophysics and Space Science, University of California San Diego, La Jolla, CA 92093, USA \\
$^{6}$Centre de Recherche Astrophysique de Lyon UMR5574, Universit{\'e} Lyon, ENS de Lyon, Universit{\'e} Lyon 1, CNRS,  F-69007, Lyon, France \\
$^{7}$Zentrumf{\"u}r Astronomie der Universit{\"a}t Heidelberg, Landessternwarte K{\"o}nigstuhl 12, D-69117 Heidelberg, Germany  \\
$^{8}$Istituto Nazionale di Astrofisica, Osservatorio Astronomico di Torino, Strada Osservatrio 20, I-10025 Pino Torinese, Italy 
}

\date{Accepted 2016 September 22. Received 2016 September 22; in original form 2016 April 5}

\begin{document}

\label{firstpage}
\pagerange{\pageref{firstpage}--\pageref{lastpage}}
\maketitle
\begin{abstract}
We have conducted a search for L subdwarf candidates within the photometric catalogues of the UKIRT Infrared Deep Sky Survey and Sloan Digital Sky Survey. Six of our candidates are confirmed as L subdwarfs spectroscopically at optical and/or near-infrared wavelengths. We also present new optical spectra of three previously  known L subdwarfs (WISEA J001450.17-083823.4, 2MASS J00412179+3547133 and ULAS J124425.75+102439.3). We examined the spectral type and metallicity classification of subclasses of known L subdwarfs. We summarized the spectroscopic properties of L subdwarfs with different spectral types and subclasses. We classify these new L subdwarfs by comparing their spectra to known L subdwarfs and L dwarf standards.  We estimate temperatures and metallicities of 22 late-type M and L subdwarfs by comparing their spectra to BT-Settl models. We find that L subdwarfs have temperatures between 1500 and 2700 K, which are higher than similar-typed L dwarfs by around 100--400 K depending on different subclasses and subtypes. We constrained the metallicity ranges of subclasses of M, L, and T subdwarfs. We also discussed the spectral-type and absolute magnitude relationships for L and T subdwarfs.
\end{abstract}




\section{Introduction}
Metal-deficient very low-mass stars (VLMS) and brown dwarfs (BDs) are primeval populations in the Galaxy's ancient halo, and represent extremes in low metallicity and old age among Galactic populations. They can reveal the fundamental interior structure physics around the substellar mass limit, and are crucial to our understanding of complex ultra-cool atmospheres and the star formation mechanisms of the early Universe. VLMS \citep[$M \loa$ 0.5 M$_{\sun}$;][]{gros74,bara95} are red dwarfs at the low-mass end of the Hertzsprung--Russell diagram's stellar main sequence. BDs are substellar objects with masses below the hydrogen burning minimum mass, which ranges from 0.075 to 0.092 M$_{\sun}$ for solar to primordial metallicities according to theoretical models \citep{bur01}. Primeval VLMS with $M \loa$ 0.1 M$_{\sun}$ and BD have subsolar metallicity and are generally referred to as ultra-cool subdwarfs (UCSDs).

VLMS and BDs are classified as M, L, T, and Y types according to spectral morphology that is dominated by temperature-dependent chemistry and thermal emission \citep{kir91,kir99,mart99,bur02,cus11}. A massive BD could be a late-type M dwarf when it is about 0.1 Gyr old, but then cools becoming a late-type L dwarf after about 10 Gyr. L subdwarfs represent the lowest mass stars with subsolar metallicity and also include massive metal-poor BDs \citep[e.g. 2MASS J05325346+8246465, referred to as 2M0532;][]{bur08b}. L subdwarfs \citep[e.g. 2M0532; ][]{bur03} exhibit characteriztic spectral signatures due to strong metal hydrides (e.g. FeH), weak or absent metal oxides (e.g. VO and CO), and enhanced collision-induced H$_2$ absorption \citep[CIA H$_2$;][]{bat52,bor89,bor01,abe12,sau12} in the near-infrared (NIR). 

Modern large-scale optical and NIR surveys have the capability to identify L subdwarfs, although they are very rare compared to L dwarfs. About 22 L subdwarfs have been reported in the literature from different surveys (see Section \ref{ssspt}). The Two Micron All Sky Survey \citep[2MASS;][]{skr06} observed in three NIR filters ($J, H$, and $Ks$), and searches therein have yielded eight L subdwarfs \citep{bur03,bur04,bur04b,bur08c,cus09,kir10}. \citet{sch04} discovered an L subdwarf by its high proper motion, measured across 2MASS and SuperCOSMOS Sky Survey epochs \citep{hamb01}.  The Sloan Digital Sky Survey \citep[SDSS; ][]{yor00} has imaged 14555 deg$^2$ of the sky in five optical bands ($u, g, r, i, z$), yielding several L subdwarfs with $i$ and $z$ band detections.  In addition two L subdwarfs have been identified using the SDSS spectroscopic survey \citep[e.g. ][]{siv09,bowl10,sch10,bur10}. The UKIRT Infrared Deep Sky Survey \citep[UKIDSS; ][]{law07} Large Area Survey (hereafter ULAS) has imaged 3500 deg$^2$ of sky in four NIR filters ($Y, J, H$, and $K$), and is about three magnitudes deeper than 2MASS (thus being sensitive to a volume of about 5.5 times larger). UKIDSS has yielded three L subdwarfs to date \citep[e.g. ][]{lod10,lod12}. Most recently the Wide-field Infrared Survey Explorer \citep[WISE;][]{wri10} has revealed eight L subdwarfs \citep{luhm14,kir14,kir16}. 

Model atmospheres \citep{all95,wit09} have been developed and used to characterize VLMS and BD \citep[e.g.][]{bur09}. The  BT-Settl  models \citep{alla11,alla13,alla14} cover a wide range of metallicity, and their success at reproducing observed L subdwarf spectral energy distributions (SEDs) suggests that they are an effective means to estimate their atmospheric parameters. 

The classification scheme for L subdwarfs has not been fully established due to the small number of confirmed objects. A  method is proposed to assign spectral types for L subdwarfs by comparing their optical spectra to those of L dwarfs \citep{bur07}. Metallicity subclasses for L subdwarfs are also unclear; however, d/sdL (mildly metal-poor), sdL, and esdL (extremely metal-poor) subclasses have been proposed \citep[e.g.][]{bur07,kir10}, and metallicity-sensitive signatures are observed in a number of L subdwarf spectra \citep[e.g. Fig 29 of][]{kir10}. 

To properly understand and characterize L subdwarfs, it is necessary to identify a sample that covers  a wide  range of effective temperature ($T_{\rm eff}$) and metallicity. In this paper we present the discovery of six new L subdwarfs. Our candidate selection process is presented in Section \ref{ssele}.  Section \ref{sspec} presents the follow up spectroscopic observations. Section \ref{sclas} describes our spectral classification and characterization of L subdwarfs. Atmospheric properties of UCSDs derived through model comparison are presented in Section \ref{smodel}. Finally, Sections \ref{sdisc} and \ref{ssumm} present further discussion and a summary. 

\begin{figure*}
\begin{center}
   \includegraphics[angle=0,width=\textwidth]{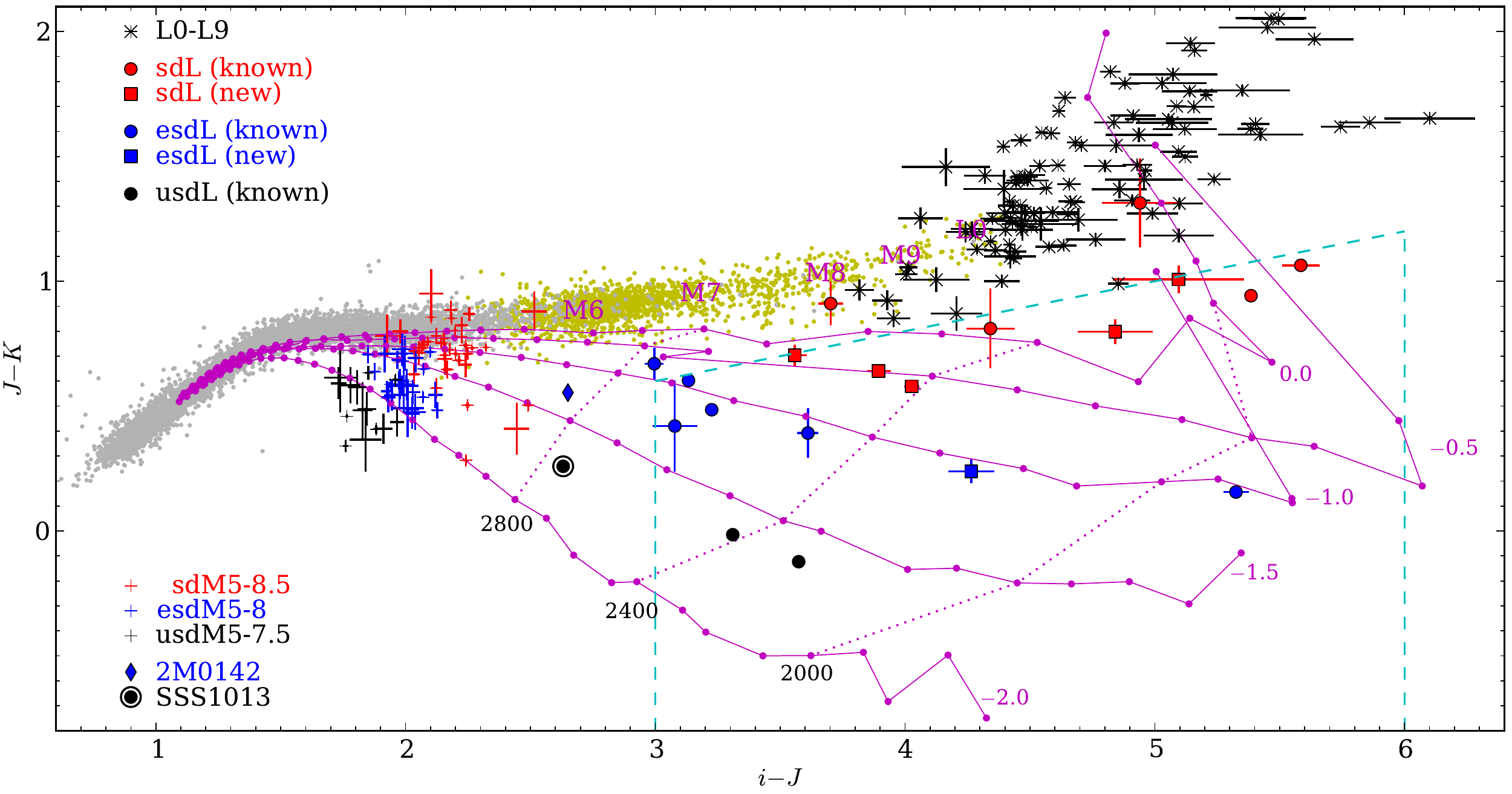}
\caption[]{The $i-J$ versus $J-K$ colours of L subdwarfs compared to M and L dwarfs.  Filled circles are 14 known L subdwarfs (with updated metallicity subclasses from this paper, red for sdL, blue for esdL, and black for usdL) from the literature with SDSS detections. Filled squares are the six new L subdwarfs (red for sdL, and blue for esdL) from this paper. Red, blue, and black crosses are sdM5-8.5, esdM5-8, and usdM5-7.5 subdwarfs confirmed with SDSS spectra and classified based on \citet{lep07}. A diamond filled with blue is 2MASS J014231.87+052327.3 \citep[2M0142;][]{bur07}. SSSPM 1013-1356 \citep[SSS1013;][]{sch04b} is indicated with a black filled circle and a larger open circle. 2MASS photometry of some known L subdwarfs has been converted into the MKO system according to \citet{hew06}. Some objects do not show error bars because these are smaller than the symbol size. Grey dots are 5000 point sources selected from a 10 deg$^2$ area of UKIDSS with $14<J<16$. Yellow dots are 1820 spectroscopically confirmed late-type M dwarfs (for which mean spectral types are indicated) from \citet{wes08}. Black asterisks are L dwarfs from DwarfArchives.org with UKIDSS and SDSS detections. The BT-Settl model grids \citep{alla14,bara15} with log $g$ = 5.5 (magenta) are over plotted for comparison, with $T_{\rm eff}$ and metallicity being indicated. The dashed cyan lines indicate our $i-J$ and $J-K$ colour selection criteria [equations (3) and (4)].}
\label{ijk}
\end{center}
\end{figure*}

\begin{table*}
 \centering
  \caption[]{Photometry of six new and five known L subdwarfs in our sample. References:  1 -- this paper; 2 -- \citet{lod12}; 3 -- \citet{kir10}; 4 -- \citet{lod12}; 5 -- \citet{bowl10} and \citet{sch10}. } 
\label{tsdlm}
  \begin{tabular}{c c c c c c c c c}
\hline
    Name  & SpT & SDSS \emph{i} & SDSS \emph{z} & UKIDSS \emph{Y} & UKIDSS \emph{J} & UKIDSS \emph{H} & UKIDSS \emph{K}    & Ref \\
\hline
ULAS J021642.97+004005.6   & sdL4~   & 22.14$\pm$0.15 & 20.03$\pm$0.10 & 18.41$\pm$0.05 & 17.30$\pm$0.03 & 16.96$\pm$0.04 & 16.51$\pm$0.04 & 1  \\
ULAS J124947.04+095019.8 & sdL1~ & 20.39$\pm$0.04 & 18.66$\pm$0.04  & 17.62$\pm$0.02 & 16.83$\pm$0.02 & 16.40$\pm$0.03 & 16.12$\pm$0.04 & 1   \\
SDSS J133348.24+273508.8  & sdL1~ & 20.51$\pm$0.05 & 18.75$\pm$0.04 & 17.47$\pm$0.02 & 17.47$\pm$0.02 & 16.62$\pm$0.01 & 16.00$\pm$0.02  & 1   \\
ULAS J133836.97$-$022910.7 & sdL7~ & 22.47$\pm$0.26 & 20.06$\pm$0.14 & 18.56$\pm$0.06 & 17.37$\pm$0.03 & 16.81$\pm$0.04 & 16.37$\pm$0.05 & 1  \\ 
SDSS J134749.74+333601.7  & sdL0~ & 19.87$\pm$0.03 & 18.06$\pm$0.02  & 16.66$\pm$0.01 & 15.85$\pm$0.01 & 15.46$\pm$0.01 & 15.27$\pm$0.02  & 1  \\
ULAS J151913.03$-$000030.0  & esdL4~~ & 21.46$\pm$0.09 & 19.33$\pm$0.06 & 18.19$\pm$0.03 & 17.21$\pm$0.02 & 17.07$\pm$0.03 & 16.97$\pm$0.04   & 1  \\
\hline
ULAS J033350.84+001406.1 & sdL0~ &19.24$\pm$0.02&17.87$\pm$0.02&16.81$\pm$0.01&16.11$\pm$0.01&15.77$\pm$0.01&15.50$\pm$0.02  & 2\\
2MASS J11582077+0435014 & sdL7~ &21.02$\pm$0.08&18.15$\pm$0.03&16.61$\pm$0.01&15.43$\pm$0.00&14.88$\pm$0.01&14.37$\pm$0.01 & 3\\
ULAS J124425.90+102441.9 & esdL0.5 &19.48$\pm$0.02&18.01$\pm$0.02&16.98$\pm$0.01&16.26$\pm$0.01&16.00$\pm$0.01&15.77$\pm$0.02 & 2 \\
ULAS J135058.86+081506.8 & usdL3 &21.25$\pm$0.08&19.52$\pm$0.06&18.66$\pm$0.05&17.93$\pm$0.04&18.07$\pm$0.10&17.95$\pm$0.15 & 4 \\
SDSS J141624.08+134826.7 & sdL7~ &18.37$\pm$0.02&15.89$\pm$0.02&14.26$\pm$0.00&12.99$\pm$0.00&12.47$\pm$0.00&12.05$\pm$0.00 & 5 \\
\hline
\end{tabular}
\end{table*}

\section{Candidate selection}
\label{ssele}
L subdwarfs are kinematically associated with the Galactic halo and thick disc, and thus they generally have high space velocities relative to the Sun, and hence have higher proper motions and larger dispersion of radial velocities than the disc population. L subdwarfs also have bluer optical and NIR colours (e.g. $i-J$ and $J-K$, see Fig. \ref{ijk}) than L dwarfs due to a variety of factors including flux suppression due to enhanced CIA H$_2$ which is stronger in the $K$ band than in the $J$ band. 
We conducted a search for L subdwarf candidates by combining the ULAS and SDSS data bases. We used both photometric and proper motion ($>$ 100 mas yr$^{-1}$) criteria to select L subdwarf candidates from the 10th data release of ULAS and the 8th data release of SDSS, which have a coverage overlap of over 3000  deg$^2$.  Our photometric selection criteria consist of five colour cuts and one magnitude cut: 
\begin{eqnarray}
Y-J > 0.6 \\
J-K < Y-J  \\
J-K < 0.2 \times (i-J) \\
3.0 < i-J < 6.0 \\
1.4 < z-J < 3.2 \\
12<J<18.2.
\end{eqnarray}
These criteria are based on the colours of known L subdwarfs \citep[e.g. table 6 in][]{kir14} and consideration of the colours of M, L, and T dwarfs \citep{wes08,day13} which we wish to reject. Criterion (1) rejects early-type stars which have bluer $Y-J$ colour. Criteria (2) and (3) reject M and L dwarfs which have redder $J-K$ colours. Criteria (4) and (5) reject M subdwarfs and  T dwarfs which are bluer and redder (by $i-J$ or $z-J$ colours) than L subdwarfs, respectively. Criterion (6) rejects bright early-type stars and targets which are too faint to have  good optical detection by SDSS or difficult for spectroscopic follow up. To take account of a broader range of SDSS imaging, we also performed a visual inspection of candidates using the SDSS Navigate tool. Known L subdwarfs all appear red in the combined $g,r,i$ false colour images presented by Navigate, and we thus rejected objects that appeared as blue, yellow or orange. Typically, such objects are mismatches or earlier-type objects with poor photometric calibration. 

Objects that survived our colour cuts and visual inspection were advanced for proper motion assessment based on ULAS and SDSS multi-epochs imaging \cite[following][]{zha09}.  
Proper motions were calculated based on coordinate and epoch differences between SDSS and UKIDSS observations. We only use proper motions for 80 percent of our candidates which have baselines of 1--10 yr. Objects with proper motion less than 100 mas yr$^{-1}$ were rejected unless they had very blue $J-K < 0.3$. We thus only used our proper motion criterion for less extreme colours where contamination rates will be greater. The proper motion criterion was not adopted for the 20 percent of objects for which the SDSS--UKIDSS baseline was less than a year. 

In this way we selected 66 candidates, which included 5 previously known L subdwarfs. Six of our new candidates were subsequently confirmed spectroscopically as L subdwarfs (see Section 3), and their $J-K$ and $i-J$ colours are plotted in Fig. \ref{ijk} which provide a comparison with other populations and models. Table \ref{tsdlm} presents the photometry of five known and six new L subdwarfs. Another 28 new subdwarfs (including 1 usdL5, 6 esdL0--esdL5, and 21 sdL0--sdT0) spectroscopically confirmed from our sample will be presented in a following paper.

\begin{figure}
\begin{center}
   \includegraphics[width=\columnwidth]{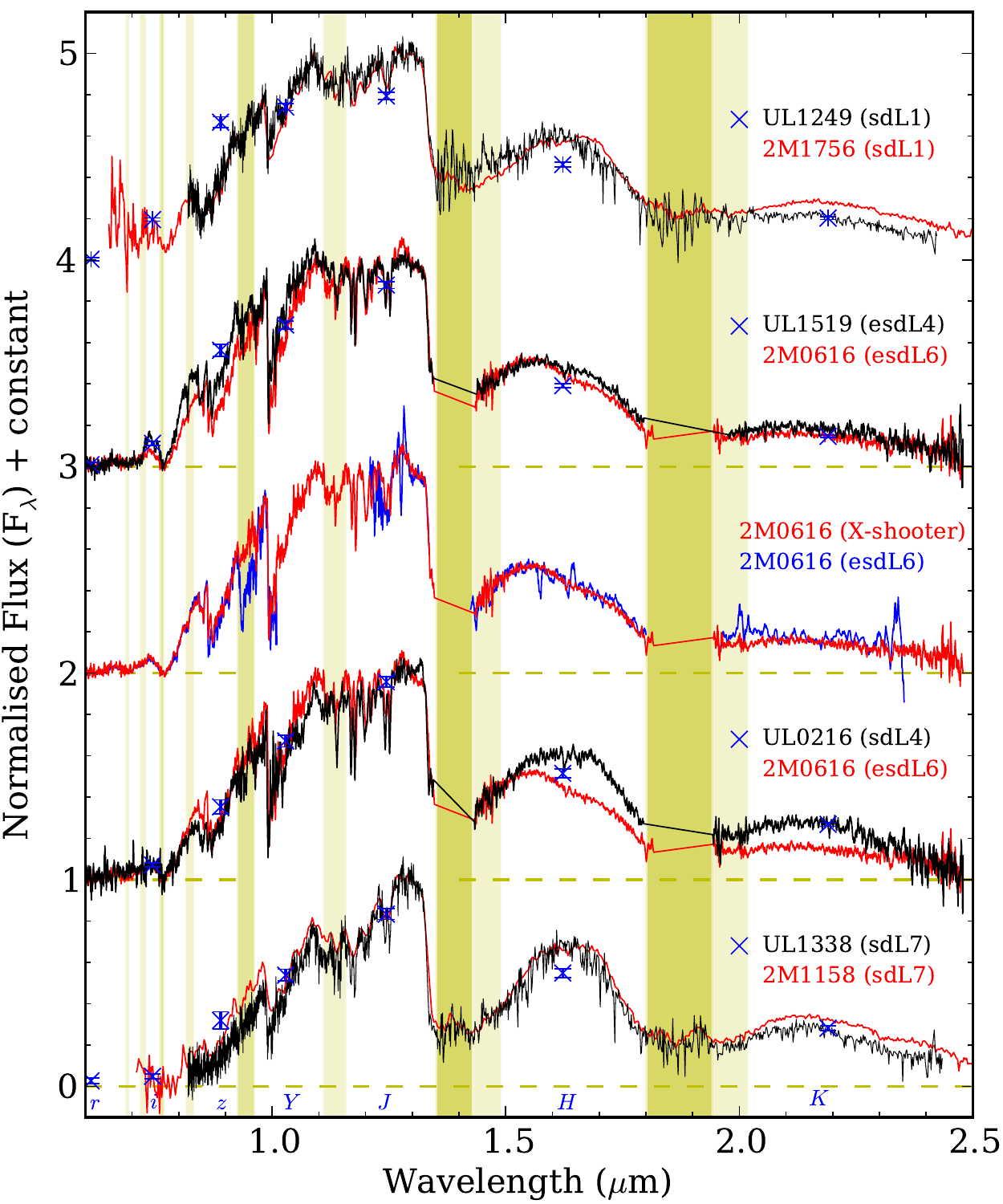}
\caption{NIR spectra of four new L subdwarfs (black) compared to known L subdwarfs (red). 
Spectra are normalised near 1.3 $\mu$m. For comparison, L subdwarf SDSS--UKIDSS photometric flux points (converted from magnitudes with VOSA; \citealt{bayo08}) are shown as blue crosses. The spectrum of 2M1756 is from \citet{kir10}. Our new X-shooter spectrum (red) of 2M0616 is over plotted with the optical and NIR spectra of 2M0616 (blue) from \citet{cus09} in the middle. Telluric absorption regions are highlighted in light yellow and have been corrected for our objects observed with X-shooter.  Light- and thick-shaded bands indicate regions with weak and strong telluric effects.}
\label{nirspec}
\end{center}
\end{figure}

\begin{figure}
\begin{center}
   \includegraphics[width=\columnwidth]{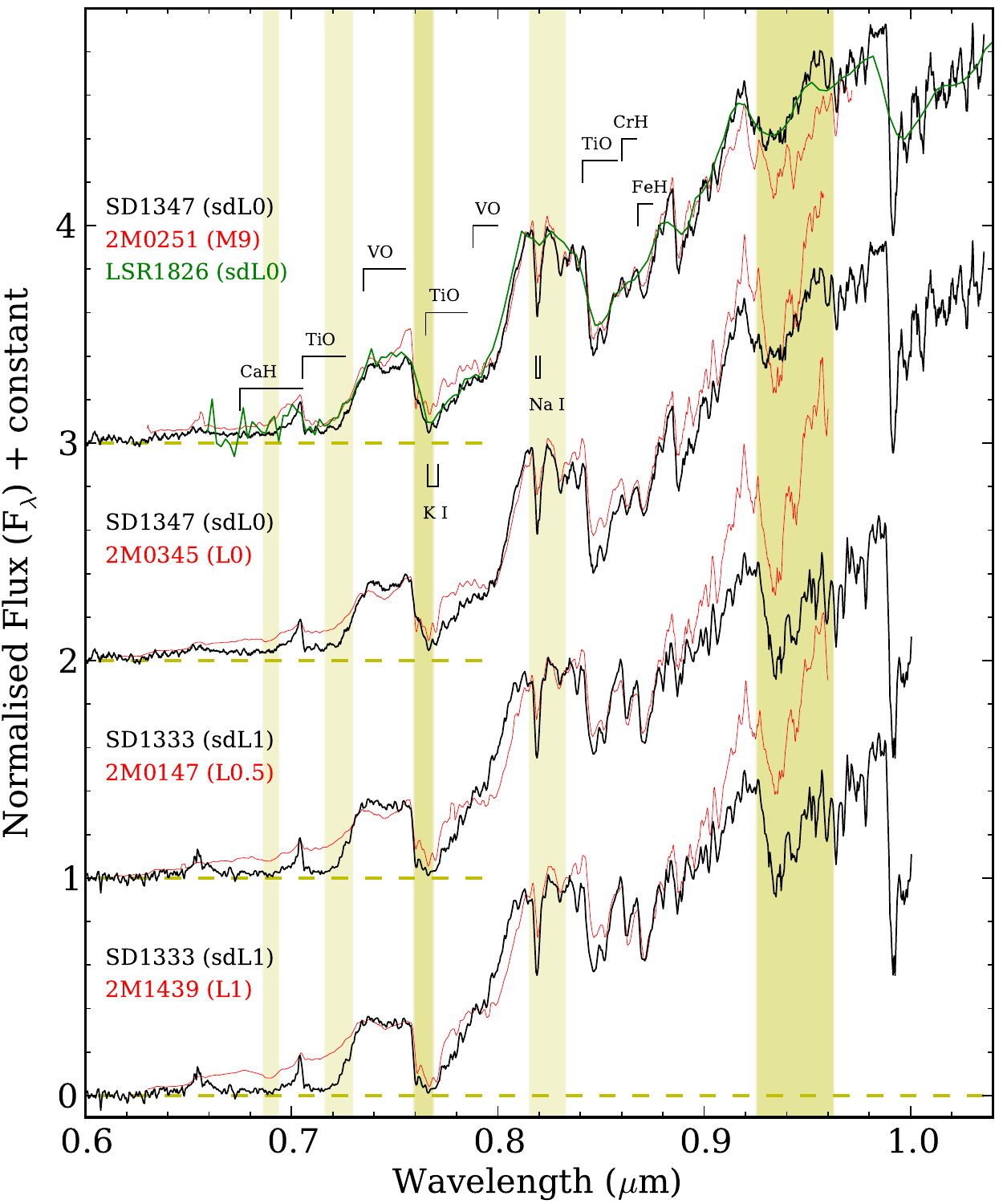}
\caption[]{Optical spectra of two new L subdwarfs (black) compared to dwarf standards (red). Spectra of 2MASS J0147334+345311 B (2M0147), 2MASS J0251222+252124 (2M0251), 2MASS J0345432+254023 (2M0345) and 2MASS J14392836+1929149 (2M1439) are from \citet{kir99}. The spectrum of LSR 1826+3014 (LSR1826) is from \citep{bur04b}. 
Spectra are normalised at 0.825 $\mu$m. Telluric absorption regions are highlighted in yellow, which are corrected for SD1347 observed by SDSS, but not corrected for SD1333 observed with OSIRIS. }
\label{2sdlop}
\end{center}
\end{figure}

\begin{landscape}
\begin{table}
 \centering
  \caption[]{Summary of the characteriztics of the spectroscopic observations.}
\label{tsdlob}
  \begin{tabular}{c c c c c c c c c c c c c  c} 
\hline
    Name   & Telescope & Instrument  & UT date & Seeing  & Airmass & $\lambda$ (VIS) & Slit & $T_{\rm int}$ & $\lambda$ (NIR) & Slit & $T_{\rm int}$  & Telluric & SpT    \\
    & & & & ($^{\prime\prime}$) & & ($\mu$m)  & (arcsec)  &  (s) & ($\mu$m)  & (arcsec)  &  (s) & Star & \\
    (1) & (2) & (3) & (4) & (5) & (6) & (7) & (8) & (9) & (10) & (11) & (12) & (13) & (14) \\
\hline
WISEA J001450.17$-$083823.4  & GTC &  OSIRIS & 2015-08-23 & 0.70 & 1.267 & 0.50--1.02 & 0.8 & ~1 $\times$ ~500 &---&---&--- & --- & --- \\
2MASS J00412179+3547133  & GTC &  OSIRIS & 2015-08-20 & 0.80 & 1.048 & 0.50--0.92 & 0.8 & ~1 $\times$ ~500 &---&---&--- & --- & --- \\
ULAS J021642.97+004005.6    & VLT & X-shooter  &  2012-01-29 & 0.67 & 1.488 &  0.53--1.02  & 0.9 & ~4 $\times$ ~400 & 0.99--2.48 & 0.9& ~4 $\times$ ~490 & HD 16031 & F0V  \\
ULAS J021642.97+004005.6    & VLT   & X-shooter  &  2014-02-17 & 0.98 & 1.252 &  0.53--1.02  & 0.9 & 12 $\times$ ~283  & 0.99--2.48 & 0.9& 12 $\times$ ~296 & HD 16031 & F0V  \\
2MASS J06164006$-$6407194  & VLT & X-shooter &  2016-01-24 & 1.19 & 1.315 & 0.53--1.02 & 1.2 & 12 $\times$ ~290   &  0.99--2.48 & 1.2 & 12 $\times$ ~300  & HR 3300 & A0V \\
ULAS J124425.75+102439.3   & Magellan & IMACS &  2010-05-05 &--- & 1.298 & 0.65--1.02 & 0.9 & ~3 $\times$ 1800  &---&---&---& --- & --- \\
ULAS J124947.04+095019.8   & Magellan & FIRE &  2012-05-08 &--- & 1.284 & ---&--- &--- & 0.82--2.50 &0.6 & ~4 $\times$ ~148 & HD110749 & A0V \\
SDSS J133348.24+273508.8  &  SDSS & SDSS & 2008-02-18 & 1.52 & 1.112 & 0.38--0.92 & 3.0 & ~1 $\times$ 2400  &---&---&---& Unknown & --- \\
SDSS J133348.24+273508.8  & GTC &  OSIRIS & 2013-12-23 & 1.10 & 1.385 & 0.50--1.00 & 1.0 & ~3 $\times$ ~900 &---&---&--- & --- & --- \\
ULAS J133836.97$-$022910.7 & Magellan & FIRE & 2012-05-08 & ---& 1.122 &--- &--- &--- & 0.82--2.50 & 0.6 & ~8 $\times$ ~148  & HD110749 & A0V \\ 
SDSS J134749.74+333601.7  & SDSS  & BOSS &  2012-10-24 & 1.37  & 1.030  & 0.36--1.04 & 2.0 & ~1 $\times$ 5405  &--- &---&---& Unknown & ---  \\
ULAS J151913.03$-$000030.0    & VLT & X-shooter  & 2012-01-29 & 1.33 & 1.819 & 0.53--1.02  & 0.9  & ~4 $\times$ ~400   & 0.99--2.48 & 0.9 & ~4 $\times$ ~490 &HD 62388 & A0V \\
ULAS J151913.03$-$000030.0   & VLT & X-shooter & 2013-04-06 & 0.63 & 1.435 & 0.53--1.02 & 0.9 & ~4 $\times$ ~205  &  0.99--2.48 & 0.9 & ~4 $\times$ ~290 & HD 130163 & A0V    \\
ULAS J151913.03$-$000030.0   & VLT & X-shooter & 2016-03-22 & --- & 1.131 & 0.53--1.02 & 1.2 & 12 $\times$ ~290  & 0.99--2.48 & 1.2 & 12 $\times$ ~300 & HR 6633 & B9/A0III \\
\hline
\end{tabular}
\end{table}
\end{landscape}

\begin{figure}
\begin{center}
   \includegraphics[width=\columnwidth]{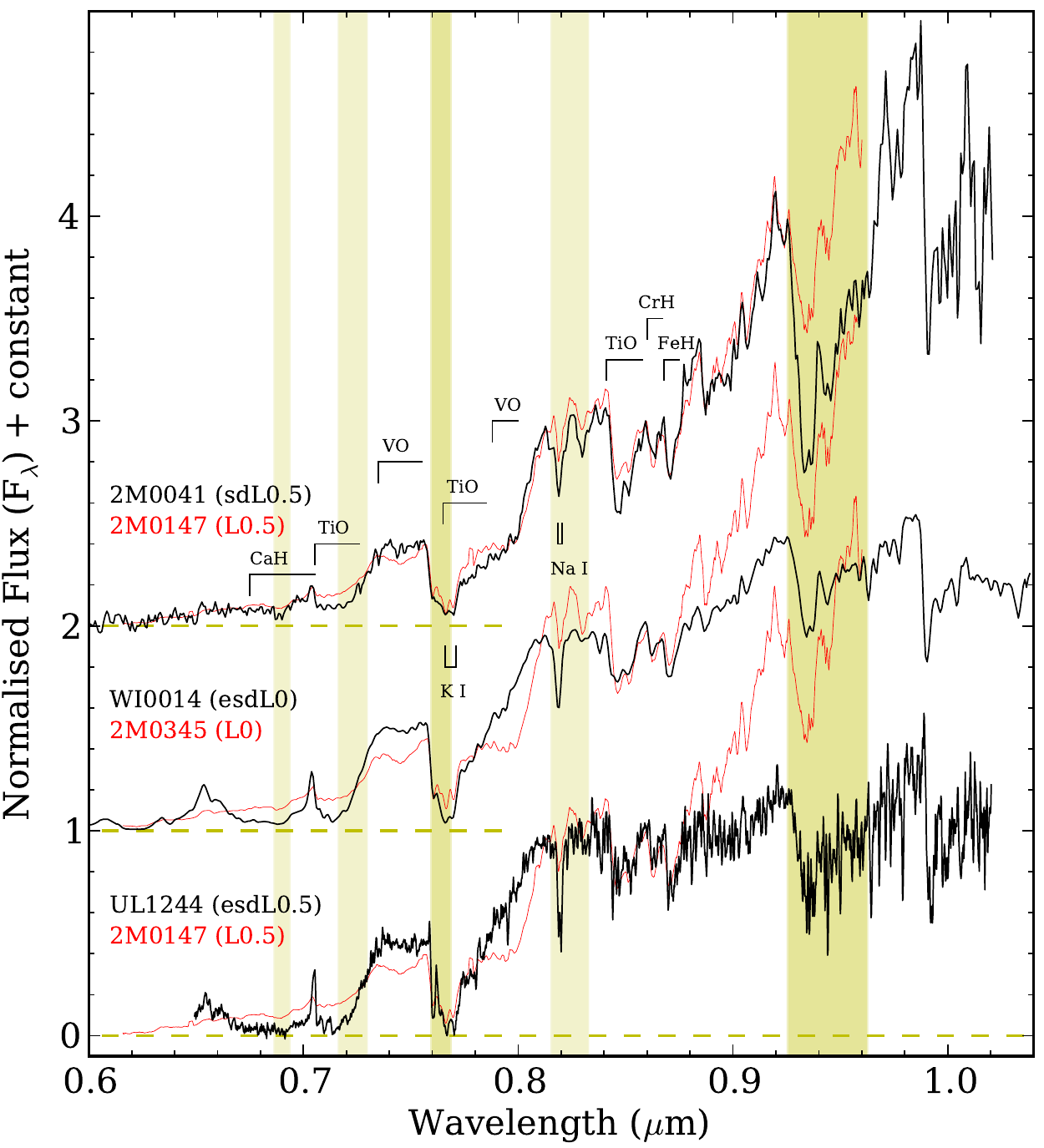}
\caption[]{New optical spectra of three known L subdwarfs (black) compared to L dwarf standards, 2M0147 and 2M0345 (red). Spectra are normalised at 0.825 $\mu$m. Telluric absorption regions are highlighted in light yellow, which are not corrected.}
\label{3sdkn}
\end{center}
\end{figure}

\section{Spectroscopic observations}
\label{sspec}
A summary of the characteriztics of the spectroscopic observations presented in this paper is given in Table \ref{tsdlob}. 
Columns 1--6  give names of targets, telescope, spectrograph, observation date, seeing and airmass. Columns 7--9 and 10--12 give wavelength ranges, slit width (fibre diameter for SDSS), numbers of exposures, and integration times for optical and NIR observations, respectively. Columns 13--14 give telluric stars and their spectral types. Observed spectra are plotted in Figs \ref{nirspec}--\ref{3sdkn}.

\subsection{New L subdwarfs}
\label{newsdlobs}

\emph{ULAS J151913.03$-$000030.0} (UL1519) and \emph{ULAS J021642.97+004005.6} (UL0216) were first confirmed with the X-shooter spectrograph \citep{ver11} on the Very Large Telescope (VLT) on 2012 January 29 with total integration times of 1960 s in the NIR and 1600 s in the visible (VIS), as backup targets of a large programme \citep{day13,maro15}. 
X-shooter has a resolving power of 5100 in the NIR arm and 8800 in the VIS arm with a 0.9 arcsec slit. With a 1.2 arcsec slit it has a resolving power of 4000 in the NIR arm and 6700 in VIS arm. A second X-shooter spectrum of UL1519 was observed in much better seeing and at lower airmass on 2013 April 6 with a total integration time of 1160 s in the NIR and 820 s in the VIS arms. We started a follow up programme of known L subdwarfs with X-shooter in 2014. We observed UL0216 on 2014 February 17 with total integration times of 3552 s in the NIR and 3396 s in the VIS. We observed UL1519 on 2016 March 22 with total integration times of 3600 s in the NIR and 3480 s in the VIS. All X-shooter spectra were observed in an ABBA nodding mode, and reduced with ESO Reflex \citep{freu13}. Telluric correction was achieved using telluric standard stars observed on the same night as our targets and at similar airmass; see Table \ref{tsdlob} for more details of our observations. 

The first and the second spectra of both UL0216 and UL1519  have a signal-to-noise ratio (SNR per pixel) of $\sim$ 2 at 0.9 $\mu$m. The first and the second spectra of UL0216 have SNR  $\sim$ 7 and $\sim$ 10  at 1.3 $\mu$m, respectively. The first and second spectra of UL1519 both have an SNR of $\sim$ 8 at 1.3 $\mu$m. The third spectrum of UL1519 has SNR of $\sim$ 12 at both 0.9 and 1.3 $\mu$m.  Two spectra of UL0216 were also combined to produce a better SNR (3 at 0.9$\mu$m and 12 at 1.3 $\mu$m) with a total integration time of 5512 s in the NIR and 4996 s in the VIS arm.  Three spectra of UL1519 were combined to produce a better SNR (13 at 0.9$\mu$m and 16 at 1.3 $\mu$m) with a total integration time of 6720 s in the NIR and 5900 s in the VIS arms. X-shooter spectra plotted in Fig. \ref{nirspec} are smoothed by 100 pixels for the VIS arm and 50 pixels for the NIR arm, which increased the SNR by a factor of 10 and 7 times, respectively and reduced the resolving power to $\sim$800 in both VIS and NIR. 

\emph{ULAS J124947.04+095019.8} (UL1249) and \emph{ULAS J133836.97-022910.7} (UL1338) were observed with the Folded-port InfraRed Echellette \citep[FIRE; ][]{simc08} spectrograph on the Magellan Telescopes on 2012 May 8, using a total integration time of 592 s for UL1249 and  1184 s for UL1338. Spectra were obtained in the prism mode which provides a resolving power of $\sim$400 near 1.25 $\mu$m. Spectra were reduced with the FIREHOSE data reduction pipeline\footnote{The pipeline tools are implemented in IDL, and are written by Rob Simcoe, John Bochanski, and Mike Matejek.  Many others have contributed unwittingly to the underlying algorithms, including Joe Hennawi, Scott Burles, David Schlegel, and Jason Prochaska. Several of the routines draw from the Spextool pipeline, written by Mike Cushing, Bill Vacca, and John Rayner.} which is based on the MASE pipeline \citep{boch09}, and the telluric correction methodology of \citet{vacc04} as integrated into SpeXtool \citep{cush03}. Telluric absorptions in UL1249 and UL1338 are corrected with an A0V star (see Table \ref{tsdlob}). Spectra of UL1249 and UL1338 have SNR of $\sim$50 and $\sim$40, respectively at around 1.3 $\mu$m. 

\emph{SDSS J133348.24+273508.8} (SD1333) and \emph{SDSS J134749.74+333601.7} (SD1347) were observed by the SDSS Legacy and BOSS spectroscopic surveys, respectively. An optical spectrum of SD1333 was observed with the original SDSS spectrographs on 2008 February 18. The SDSS spectrum of SD1333 has an SNR of about 3 at 0.9 $\mu$m. Another optical spectrum of SD1333 was obtained with the Optical System for Imaging and low-Intermediate-Resolution Integrated Spectroscopy \citep[OSIRIS;][]{cepa00} instrument on the Gran Telescopio Canaries (GTC). The spectrum was reduced using standard procedures within IRAF\footnote{IRAF is distributed by the National Optical Observatory, which is operated by the Association of Universities for Research in Astronomy, Inc., under contract with the National Science Foundation.}. It has a mean resolving power of $\sim$500 and an SNR of $\sim$150 at 0.81 $\mu$m. A B1-type star, Hilt 600, was used as a standard for flux calibration.  Telluric absorptions in the spectrum are not corrected.  An optical spectrum of SD1347 was observed with the BOSS Spectrographs on 2012 October 24. The SDSS spectrum of SD1347 has an SNR of $\sim$24 at 0.9 $\mu$m and a resolving power of $\sim$2000. Telluric absorptions in SDSS spectra are corrected.  The spectrum of SD1347 in Fig. \ref{2sdlop} is smoothed by 5 pixels for display.

\subsection{Known L subdwarfs}

\emph{ULAS J124425.75+102439.3} (UL1244) was discovered as an sdL0.5 subdwarf by \citet{lod12}. We observed it as an L subdwarf candidate with the Inamori Magellan Areal Camera and Spectrograph \citep[IMACS; ][]{dre11} Short-Camera on the Baade Magellan Telescope with a total integration time of 5400 s on 2010 May 5. The spectrum covered a wavelength range of 0.65--1.02 $\mu$m, and has a resolving power of $\sim$1000. The spectrum was reduced using standard procedures within IRAF and has an SNR of $\sim$60 around 0.81 $\mu$m. A B9V-type star, Hip 77673, was used as a standard for flux calibration. Telluric absorptions in the spectrum are not corrected. 

\emph{WISEA J001450.17-083823.4} (WI0014) was discovered as an sdL0 subdwarf in the optical \citep{kir14} and NIR \citep{luhm14}. The confirmed optical spectrum of WI0014 has a spectral range covering 0.55--0.8 $\mu$m, and we obtained a new optical spectrum covering the 0.65--1.02 $\mu$m with OSIRIS on 2015 August 23. The OSIRIS spectrum of WI0014 has a resolving power of $\sim$300, and an SNR of $\sim$300 at 0.81 $\mu$m. \emph{2MASS J00412179+3547133} (2M0041) was identified as an sdL candidate by its NIR spectrum \citep{bur04b}. There are no optical spectra of 2M0041 in the literature, and we Therefore, obtained an optical spectrum with  OSIRIS on 2015 August 20. The OSIRIS spectrum of 2M0041 has a mean resolving power of $\sim$300, and an SNR of $\sim$70 at 0.81 $\mu$m. Spectra of WI0014 and 2M0041 were reduced using standard procedures within IRAF. A DZA5.5-type white dwarf, Ross 640, was used as a standard for flux calibration. Telluric absorptions in the spectrum are not corrected.

\emph{2MASS J06164006$-$6407194} (2M0616) was discovered by \citep{cus09} with optical and NIR spectra observed individually. The 1.0--1.2 $\mu$m spectrum of 2M0616 is missing. We observed 2M0616 with X-shooter on 2016 January 24. The total integration time is 3600 s in the NIR and 3480 s in the VIS. The observation and data reduction are performed in the same way as UL1519 (see Section \ref{newsdlobs}). The spectrum of 2M0616 has SNR of $\sim$ 15 at 0.9 $\mu$m and $\sim$ 18 at 1.3 $\mu$m.

\section{Classification \& characterization}
\label{sclas}

The classification of UCSDs is a challenge for several reasons. First, a wide variety of both optical and NIR spectral features are sensitive not only to $T_{\rm eff}$ changes, but also to a wide range of metallicities. Secondly, the sample of known UCSDs (particularly L type) is small. And thirdly, there are no well-resolved UCSD companions (to the more common subdwarf stars) that can be used to calibrate the metallicity consistency of a classification scheme. 

\subsection{Classification schemes for ultra-cool subdwarfs}
\label{ssmsd}

\citet{bur07} extended the M subdwarf classification scheme of \citet{giz97} out into the late M- and L-type regimes. \citet{giz97} tested the spectroscopic metallicity scale of their subclasses of M subdwarfs with $Hubble Space Telescope$ photometry of globular clusters, but this test was only done for early M spectral types. \citet{kir16} proposed a spectral sequence of late-type M and L subdwarfs as an extension of the M subdwarf classification scheme of \citet[][hereafter LRS07]{lep07}. LRS07 used a metallicity index $\zeta_{\rm TiO/CaH}$ to define metallicity subclasses of M subdwarfs. The $\zeta_{\rm TiO/CaH}$ index is based on CaH2, CaH3 and TiO5 indices, which are calculated from the ratio of the average flux over 6814--6846 \AA~(CaH2), 6960--6990~\AA~(CaH3), 7126--7135 \AA~(TiO5), and 7042--7046\AA~(Denominator), see table 1 of LRS07). The consistency of $\zeta_{\rm TiO/CaH}$ as a metallicity index was examined using six resolved binaries (whose components would be expected to share the same metallicity) containing early-type M subdwarfs. The metallicity consistency of subclasses of mid--late types (e.g. sdM3+ and esdM5+) could not be tested due to the lack of  binaries with companions in this spectral type/subclass domain.

Fig. \ref{ijk} shows four objects lying between the esdM5--esdM8 subdwarfs and SSS1013 \citep[which has been classified as esdM9.5 by][]{bur07}, but classified as late-type sdM according to LRS07. This means late-type sdMs classified according to LRS07 could be as metal-poor as mid type esdMs. This is because the metallicity is not consistent across all M subtypes defined by LRS07. The metallicity consistency is tested only for early-type M subdwarfs (<esdM3.5 and <usdM6) in their Fig. 6.  The NextGen models \citep{hau99} supported the metallicity consistency of subclasses for early-type esdM and usdM subdwarfs, but not for the late-types. Fig. 8 of LRS07 shows the isometallicity data points derived from the NextGen model grid and their metallicity subclass boundaries in a space of CaH2+CaH3 versus TiO5. These isometallicity data points with log $Z$ = --1.0 and --2.0 fit in between the sdM--esdM and esdM--usdM boundaries at CaH2+CaH3 > 1.0 (equivalent to esdM3.5 or usdM3.5). Then these isometallicity data points start to go off the middle of the subclass boundaries, and finally cross these boundaries at around CaH2+CaH3 = 0.5 (equivalent to esdM7.5 or usdM7.5). The solar metallicity model data points do not follow the M dwarf sequence in fig. 8 of LRS07, presumably because M dwarfs have more complicated atmospheres and are more difficult to reproduce with models compared to M subdwarfs. 

\begin{figure}
\begin{center}
   \includegraphics[width=\columnwidth]{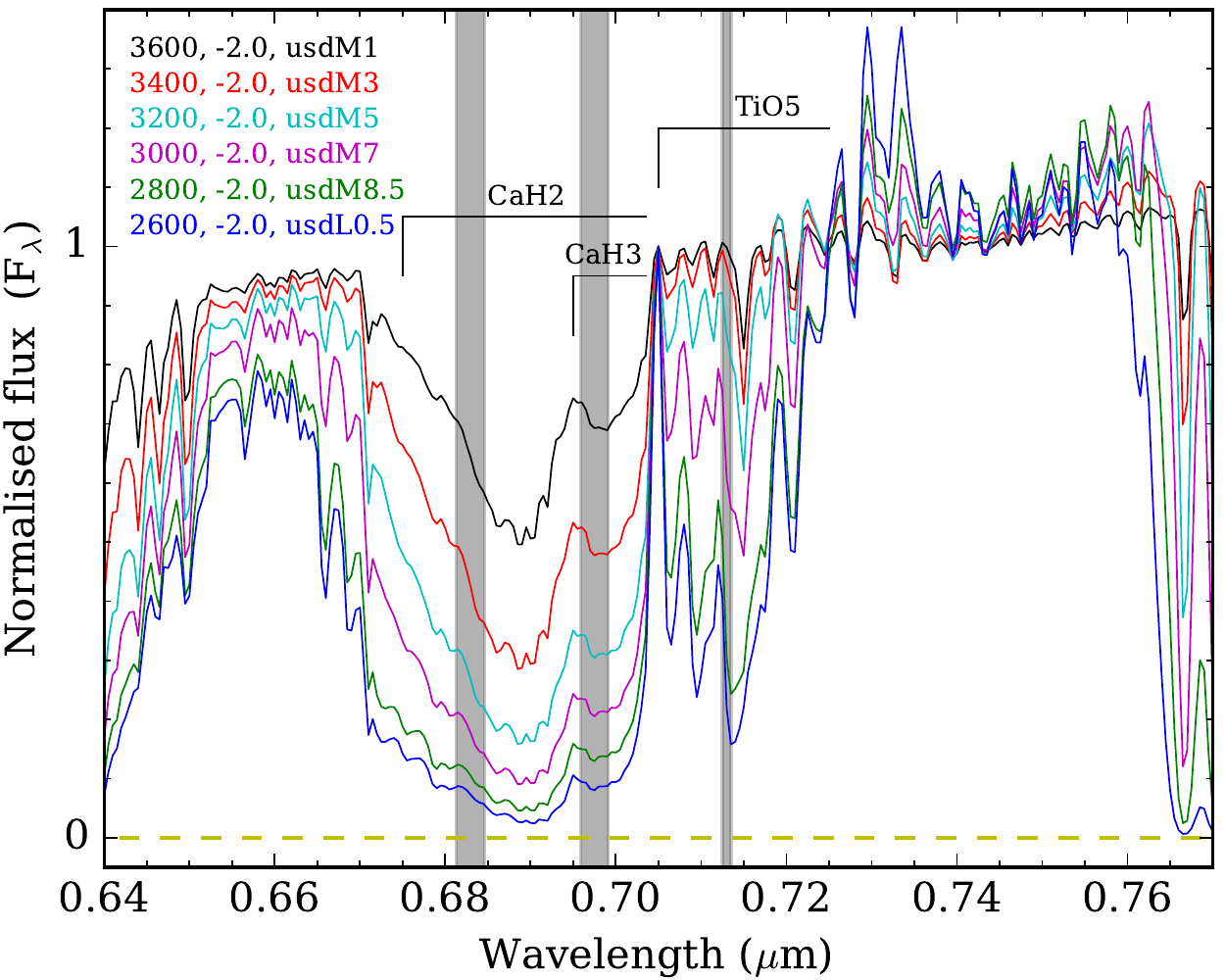}
\caption[]{BT-Settl model spectra with an [Fe/H] of --2.0 and $T_{\rm eff}$ of 3600, 3400, 3200, 3000, 2800 and 2600 K \citep{alla11}. Approximate spectral types of each model spectrum are given based on model fitting of optical spectra of known UCSDs. Spectral wavelengths shaded in grey are regions used to define CaH2 (0.6814--0.6846 $\mu$m), CaH3 (0.6960--0.6990 $\mu$m) and TiO5 (0.7126--0.7135 $\mu$m) indices. Spectra are normalised to an average of unity over the range 0.7042--0.7246 $\mu$m. Thus the average fluxes of the shaded areas represent the strengths of the CaH2, CaH3 and TiO5 indices. Equivalent spectral types of these model spectra are based on model fitting of optical spectra of M subdwarfs.}
\label{cahtiot}
\end{center}
\end{figure}

\begin{figure}
\begin{center}
   \includegraphics[width=\columnwidth]{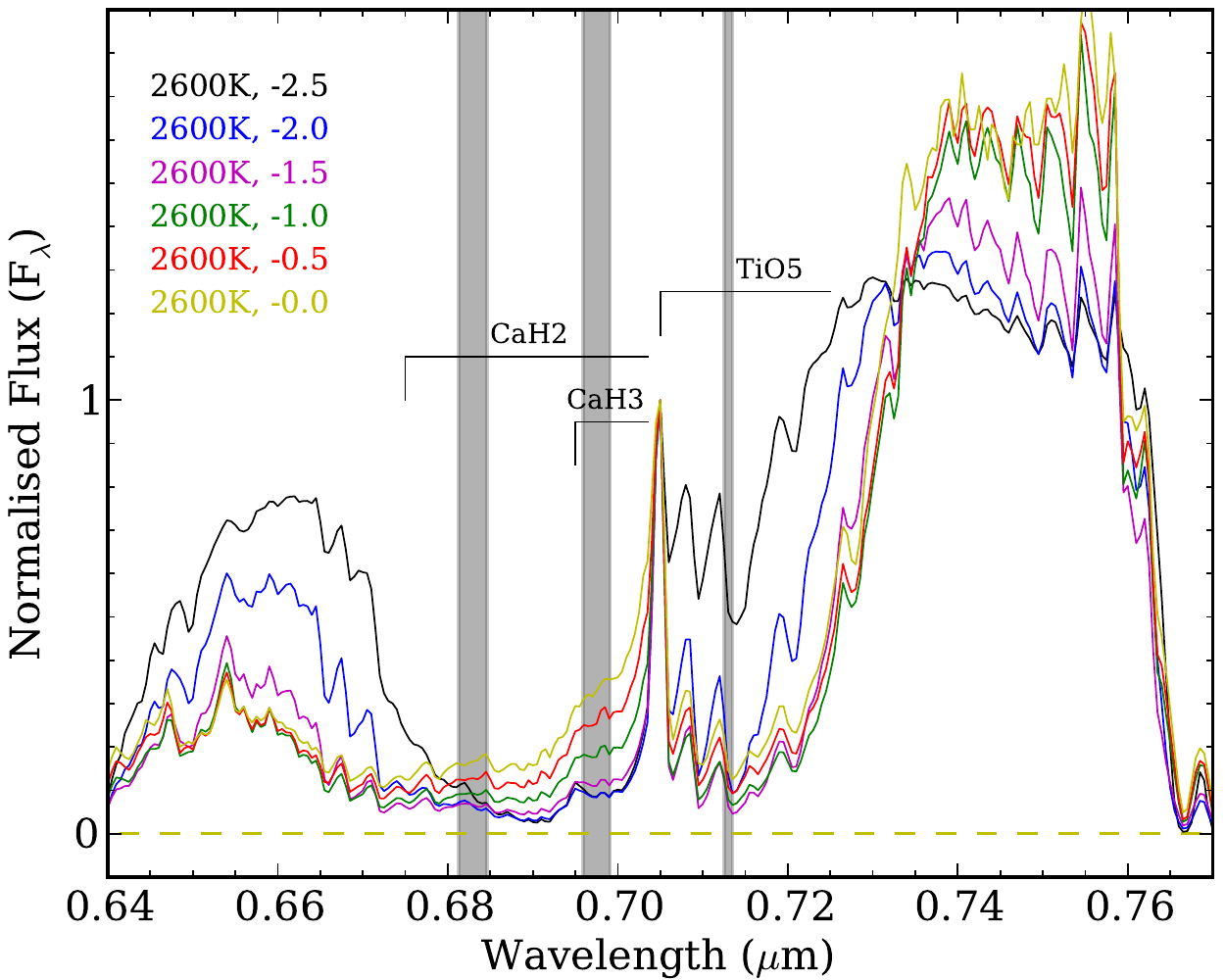}
\caption[]{BT-Settl model spectra with an $T_{\rm eff}$ of 2600 K, and [Fe/H] of 0.0, --0.5, --1.0, --1.5, --2.0 and --2.5 \citep{alla14}.  Shaded areas are explained in the caption to Fig. \ref{cahtiot}. }
\label{cahtiom}
\end{center}
\end{figure}

The TiO5 band becomes more sensitive to temperature than metallicity for late-type M subdwarfs. Fig. \ref{cahtiot} shows that the CaH absorption bands strengthen with decreasing $T_{\rm eff}$ while the TiO5 band generally remains constant through 3600-3200 K. Then the strengthening of CaH absorption bands slows down and reaches a maximum at 2600 K, being less sensitive to temperature. However, TiO5 absorption band starts to strengthen fast after 3200 K, and becomes very strong at 2600 K. It is thus not a uniform metallicity indicator across all M subtypes. Fig. \ref{cahtiom} shows that at 2600 K, the TiO5 absorption band strengthens slowly as [Fe/H] decreases from 0.0 to --1.5, but weakens as [Fe/H] decreases from --1.5 to --2.5. The relationship between the strengths of TiO5 absorption and [Fe/H] is thus not monotonic for late-type M and early-type L subdwarfs. Simple index-based classification (e.g. using TiO and CaH) can Therefore, misrepresent [Fe/H] for later type M subdwarfs. Instead we determine subclasses via an empirical assessment of a broader range of spectral features in the optical and NIR (e.g. 0.8 $\mu$m VO and 2.3 $\mu$m CO). However, since the metallicity consistency of early M subclasses has been tested (by LRS07), we aim to anchor our classification scheme within this framework. We use classes d, sd, esd, and usd, and later show (see Section \ref{smetal}) that the metallicity ranges of these subclasses are reasonably consistent with those of the early M subdwarfs.

\subsection{Spectral classification of L subdwarfs}
\label{ssclass}

\begin{figure}
\begin{center}
   \includegraphics[width=\columnwidth]{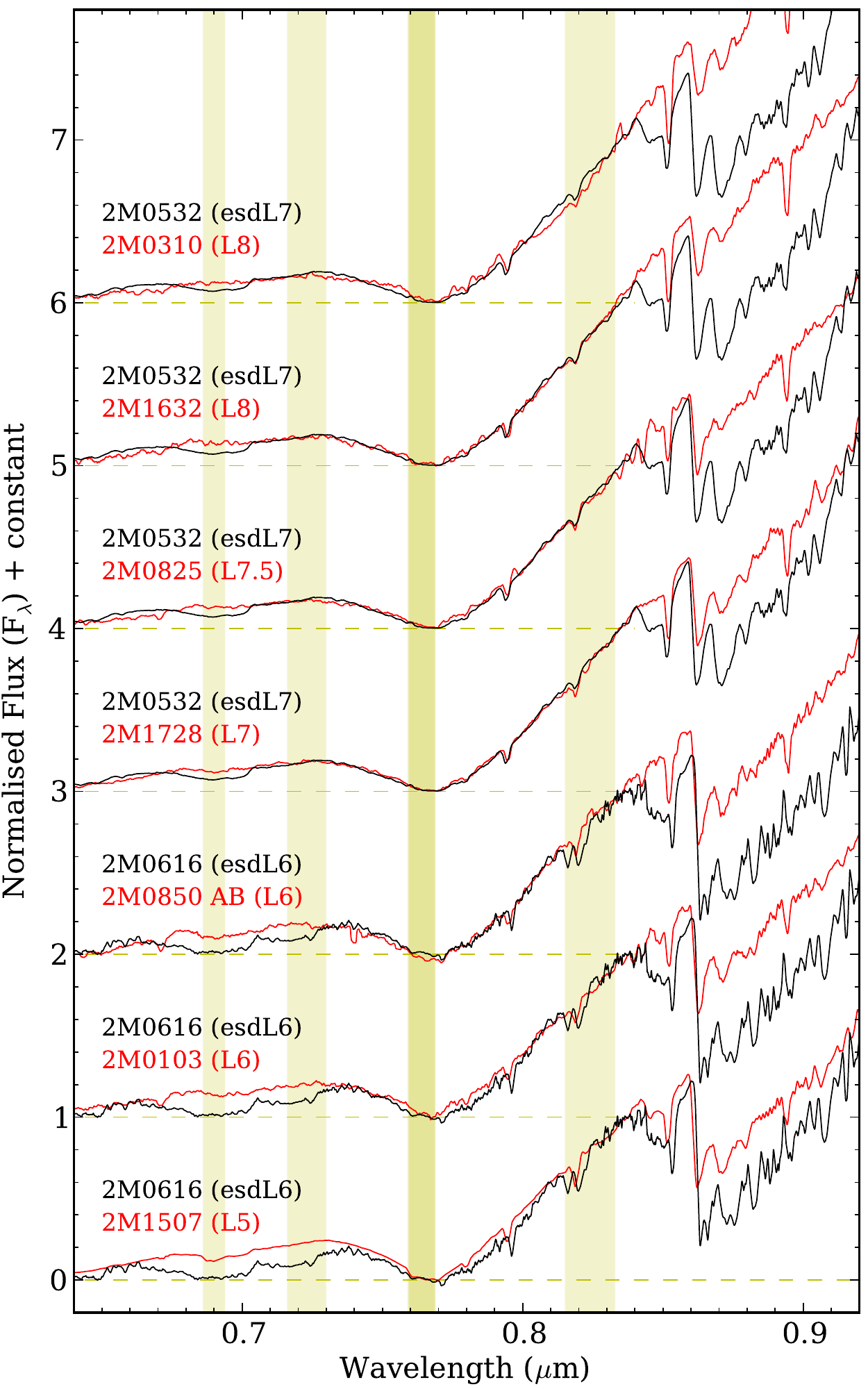}
\caption{
Optical spectra of 2M0532 \citep{bur03} and 2M0616 \citep{cus09} compared to L dwarf standards. Spectra are normalised at 0.835 $\mu$m.  The spectra of  2MASS J16322911+1904407 (2M1632) and 2MASS J08503593+1057156  AB (2M0850 AB) are from \citet{kir99}. Spectra of  2MASS J03105986+1648155 (2M0310), 2MASS J17281150+3948593 (2M1728), 2MASS J01033203+1935361 (2M0103), and 2MASS J15074769-1627386 (2M1507) are from \citet{kir00}. 
}
\label{j0616}
\end{center}
\end{figure}

Spectral types of L subdwarfs are determined by comparing their red optical spectra to those of L dwarf spectral standards \citep{bur07,kir99,kir10}. The optical spectra of L subdwarfs and dwarfs are different but comparable. We are mainly considering the 0.73--0.88 $\mu$m region to make a comparison, because this region changes constantly with type \citep[e.g.,][]{kir99} and similar features are present in the spectra of both L dwarfs and L subdwarfs. A subclass `sdL' is used to classify L subdwarfs following the `sdM' subclass of M subdwarfs \citep{giz97}. A subclass `esdL' was proposed for L subdwarfs with very strong metal-poor features \citep[e.g. 2M0532;][]{kir10}. Some marginal cases are classed as d/sdL (mildly metal-poor) if their spectra have weaker metal-poor features. \citet{bur07} defined a d/sdM subclass for late-type M subdwarfs. A d/sdL7 spectral type was used for SDSS J141624.08+134826.7 \citep[SD1416;][]{bur10}, which was classified as sdL7 by \citet{kir10}. For the naming of L subclasses, we followed the basis of LRS07 and \citet{kir10}, in which d/sdL and sdL of \citet{bur07} are generally equivalent to sdL and esdL, respectively. We also defined a usdL subclass following the suppression strength of NIR spectra caused by enhanced CIA H$_2$. 
Here we re-examine the spectral type and metallicity subclasses of some previously defined L subdwarfs, and then use them as spectral standards to classify our new sample. 

The K {\scriptsize I} doublet around 0.77 $\mu$m is one of the most notable features in the spectra of L dwarfs.  This feature is sensitive to $T_{\rm eff}$ and mildly to gravity, keeps broadening from early to late L type, and is one of the main criteria for classifying L dwarfs \citep[e.g.][]{kir99}. The first known L subdwarf, 2M0532 was classified as sdL7$\pm$1 because its optical spectrum compares well to those of L7 dwarfs \citep{bur03}. However, \citet{kir10} propose to classify 2M0532 as an esdL7 to indicate its extreme nature and unusual spectral morphology, and also suggest that 2M0532 may be somewhat later than L7. Fig. \ref{j0616} shows that 2M0532 compares well with either L7 or L7.5 spectra in the optical. 2M0532 also compares well with the L8 dwarf 2MASS J16322911+1904407 \citep{kir99}, but compares slightly less well with another L8 dwarf, 2MASS J03105986+1648155 \citep{kir00}. Although 2M0532 compares well with either L7 or L7.5 dwarfs, we suggest to classify it as esdL7 in the absence of an object with spectral features intermediate between 2M0532 and 2M0616. 2M0616 was found and classified as sdL5 by \citet{cus09}. However, Fig. \ref{j0616} shows that 2M0616 compares rather more favourably with the L6 spectral standard (compared to the L5) in the K {\scriptsize I} region. 
We thus adopt a classification of esdL6 for 2M0616 here.

\begin{figure}
\begin{center}
   \includegraphics[width=\columnwidth]{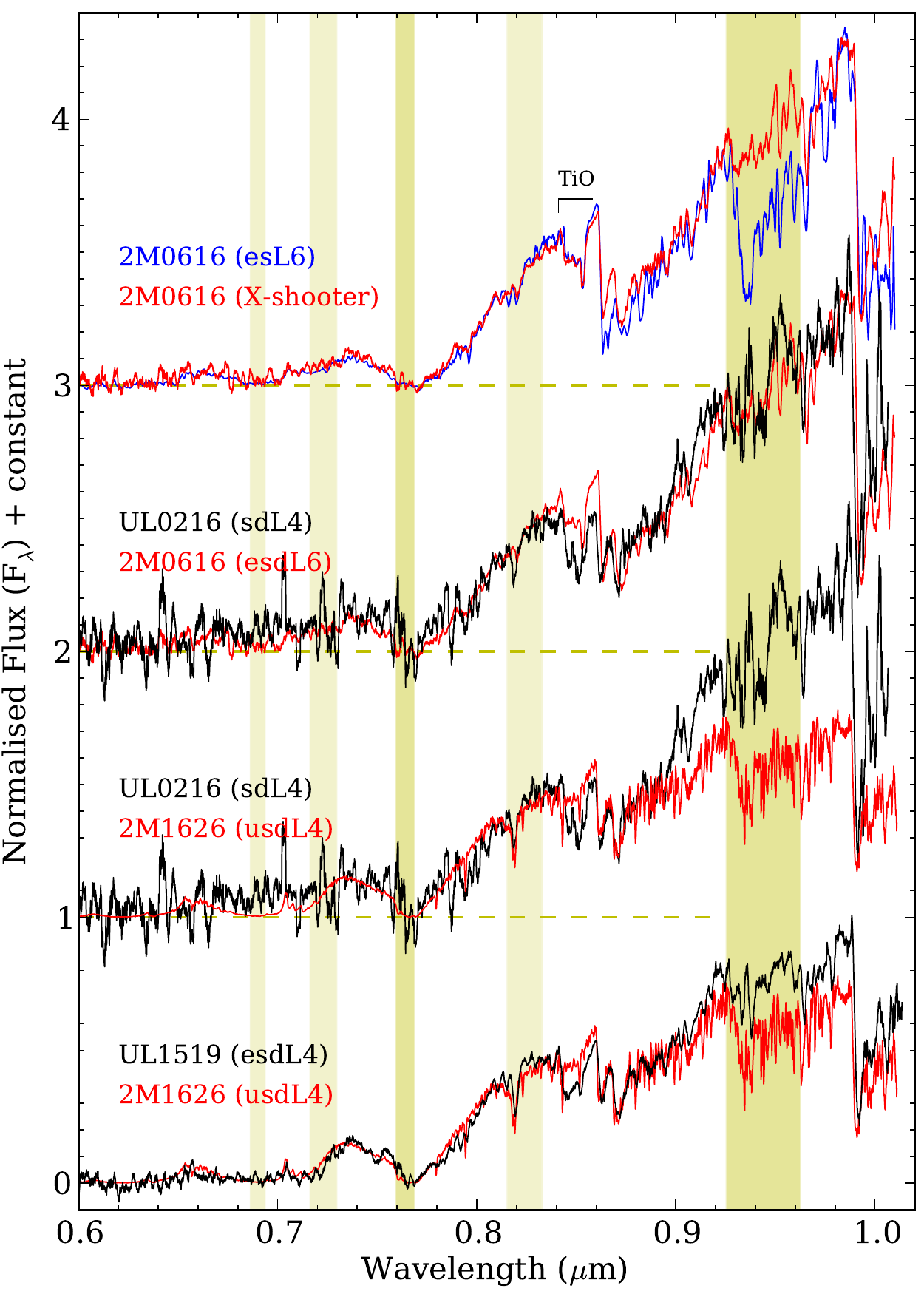}
\caption{X-shooter optical spectra of UL1519 and UL0216 compared to those of 2M1626 and 2M0616. Spectra are normalised at 0.83 $\mu$m. Telluric absorption regions are highlighted in yellow, and have been corrected for our objects observed with X-shooter.}
\label{visspec}
\end{center}
\end{figure}

Fig. \ref{visspec} shows optical spectra of UL1519 and UL0216 compared to those of 2MASS J16262034+3925190 \citep[2M1626;][]{bur04} and 2M0616. The new spectrum of 2M0616 (observed with X-shooter) compares well with the optical spectrum from \citet{cus09}, except for the telluric absorption region around 0.94 $\mu$m. UL1519 compares well with 2M1626 at 0.6-0.92 $\mu$m. Stronger TiO absorption at 0.85 $\mu$m (TiO absorption decreases from [Fe/H] = --1.5 to --2.5; Fig. \ref{cahtiom}) and extra flux beyond 0.92 $\mu$m compared to 2M1626 indicate a higher metallicity. UL0216 compares better with 2M1626 than 2M0616 at 0.6-0.89 $\mu$m. UL0216 has a higher metallicity than 2M1626 and 2M0616 because it has stronger TiO absorption at 0.85 $\mu$m and a redder spectrum than 2M1626 beyond 0.9 $\mu$m. Red optical and NIR spectra redden with increasing metallicity, and become bluer with increasing temperature. Therefore, UL0216 could have a similar spectral profile to 2M0616 at 0.9-1.0 $\mu$m, while their NIR spectra are different due to CIA H$_2$.

The CIA H$_2$ and 2.3 $\mu$m CO absorption bands are strong indicators of metallicity for L dwarfs and subdwarfs. NIR spectral emission becomes more suppressed at lower metallicity due to enhanced CIA H$_2$. The CO band is present in the spectra of late-type M, L, and early-type T dwarfs \citep[e.g.][]{kir10}. The CO band weakens as metallicity decreases, and eventually disappears.

\begin{figure*}
\begin{center}
   \includegraphics[width=\textwidth]{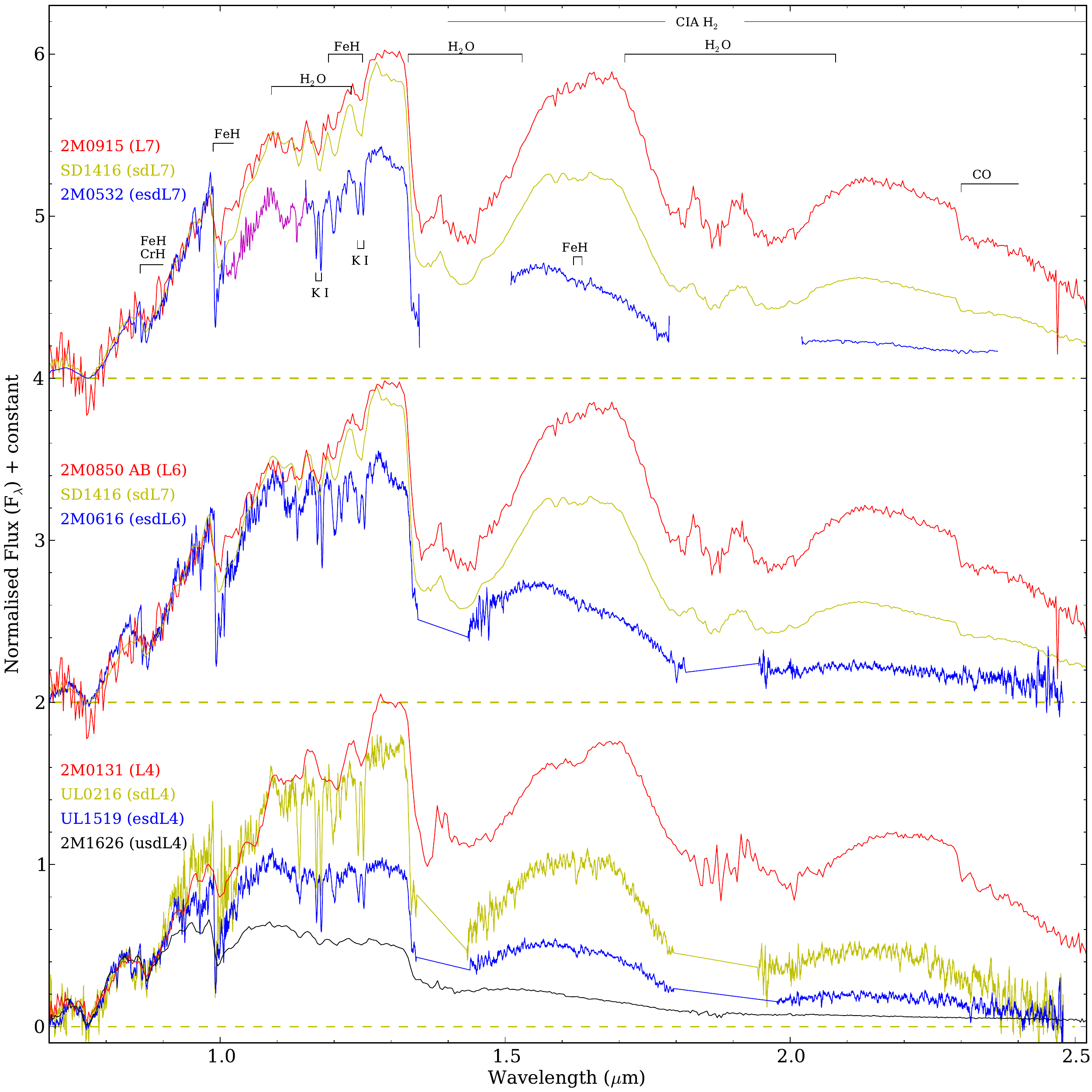}
\caption[]{
Optical and NIR spectra of L4, L6, and L7 dwarfs/subdwarfs with different sub-classes. Spectra have been normalised at 0.89 $\mu$m. The spectrum of 2M0532 at 1.008--1.153 $\mu$m wavelength is missing. The best-fitting BT-Settl model spectrum of 2M0532 ($T_{\rm eff}$ = 1600 K, [Fe/H] = --1.6, and log $g$ = 5.25) is plotted to fill the gap (in magenta).} 
\label{l7yn}
\end{center}
\end{figure*}

Fig. \ref{l7yn} shows the optical and NIR spectra of L4, L6, and L7 dwarfs and subdwarfs normalised in the optical. The top panel of Fig. \ref{l7yn} shows the spectra of 2MASS J09153413+0422045 \citep[2M0915;][]{bur07b}, SD1416 \citep[][]{sch10}, and 2M0532. Although these objects have very similar optical spectra, they show large diversity in the NIR due to very different metallicity, and can be naturally classified in three different subclasses: L7, sdL7, and esdL7. The 2.3 $\mu$m CO band gets weaker from L7 to sdL7, and disappears for esdL7. The middle panel of Fig. \ref{l7yn} shows the spectra of 2M0850 AB, 2M0616 and SD1416, which are very similar in the optical but very different in the NIR. There is no sdL6 currently known, so we show the spectra of SD1416 instead. The bottom panel of Fig. \ref{l7yn} shows the spectra of 2MASS J01311838+3801554 \citep[2M0131;][]{burg10}, UL0216, UL1519, and 2M1626.

We classified UL0216 as sdL4 because it compares well with 2M1626 at 0.6--0.89 $\mu$m (Fig. \ref{visspec}), and has a suppressed NIR spectrum due to enhanced CIA H$_2$. UL1519 compares well with 2M1626 at 0.6--0.89 $\mu$m (Fig. \ref{visspec}), and has stronger NIR suppression than UL0216, which is very similar to 2M0616  (Fig. \ref{nirspec}). Therefore, we classify UL1519 as an esdL4 subdwarf. 
2M1626 was previously classified as sdL4 based on the similarity of its optical spectrum to those of L4 dwarfs \citep{bur07b}. However, it has weaker TiO absorption at 0.85 $\mu$m (Fig. \ref{visspec}) and stronger NIR suppression compared to UL1519 suggesting it should be in a lower metallicity subclass. Therefore, we classify 2M1626 as an usdL4 subdwarf. 
2MASS J17561080+2815238 (2M1756) and 2MASS J11582077+0435014 (2M1158) are classified as sdL1 and sdL7 based on their similar optical spectra to sdL1 and sdL7 subdwarfs \citep{kir10}. Fig. \ref{nirspec} shows that the NIR spectra of UL1249 and UL1338 compare well with 2M1756 and 2M1158; thus, we classify them as sdL1 and sdL7, respectively.

\begin{figure}
\begin{center}
   \includegraphics[width=\columnwidth]{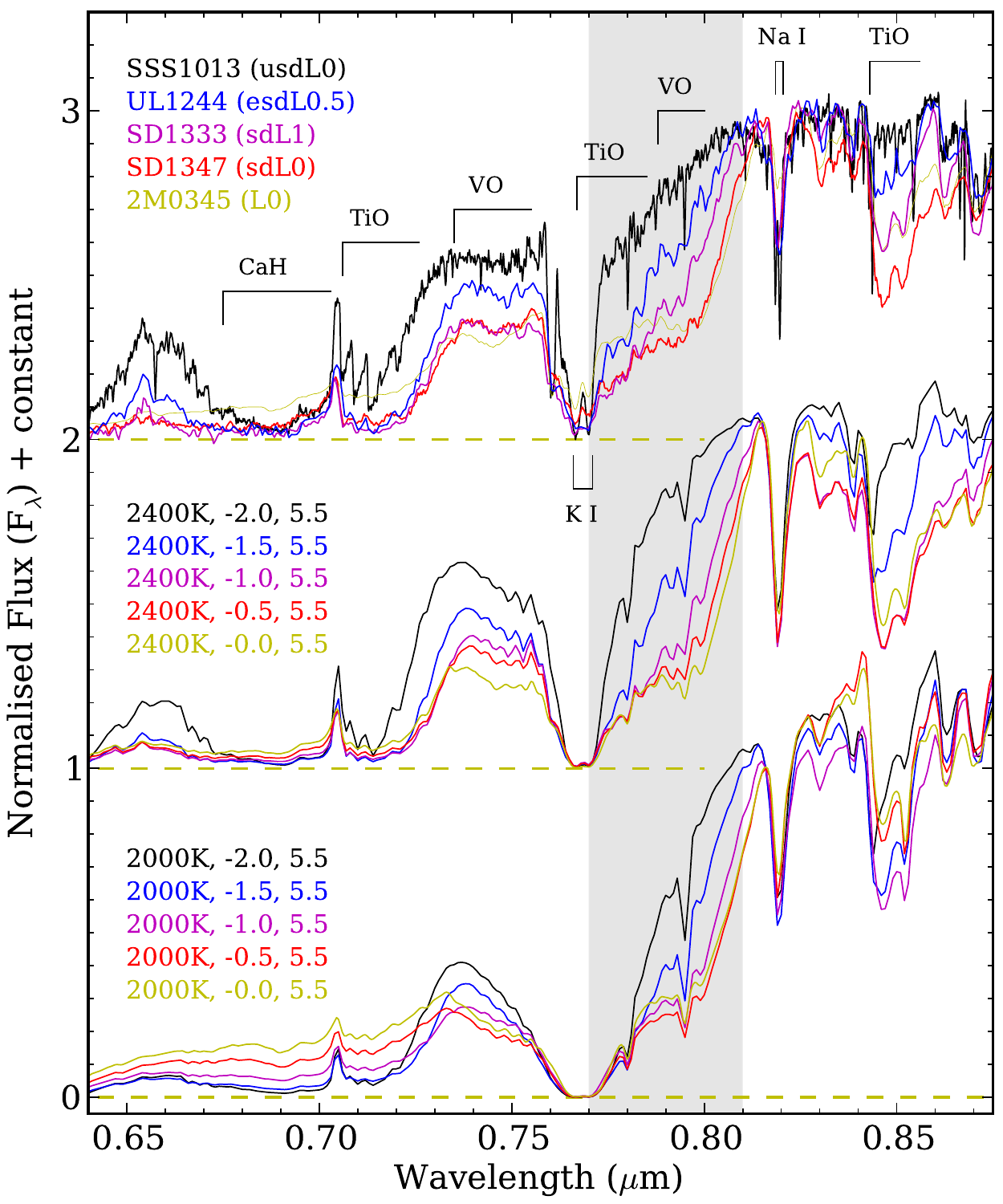}
\caption[]{Comparison of BT-Settl optical spectra with different [Fe/H] (0.0, --0.5, --1.0, --1.5, --2.0) at $T_{\rm eff}$ of 2400 and 2000 K. All spectra have $\log{g}$ of 5.5 dex. Spectra are normalised at 0.815 $\mu$m. Shaded grey area indicates the region with VO and TiO absorptions, which shows large differences between the spectra with different metallicity. Observed spectra of a few objects with similar profiles as model spectra in the middle panel are plotted on the top panel for comparison.}
\label{mspec}
\end{center}
\end{figure}

\begin{figure}
\begin{center}
   \includegraphics[width=\columnwidth]{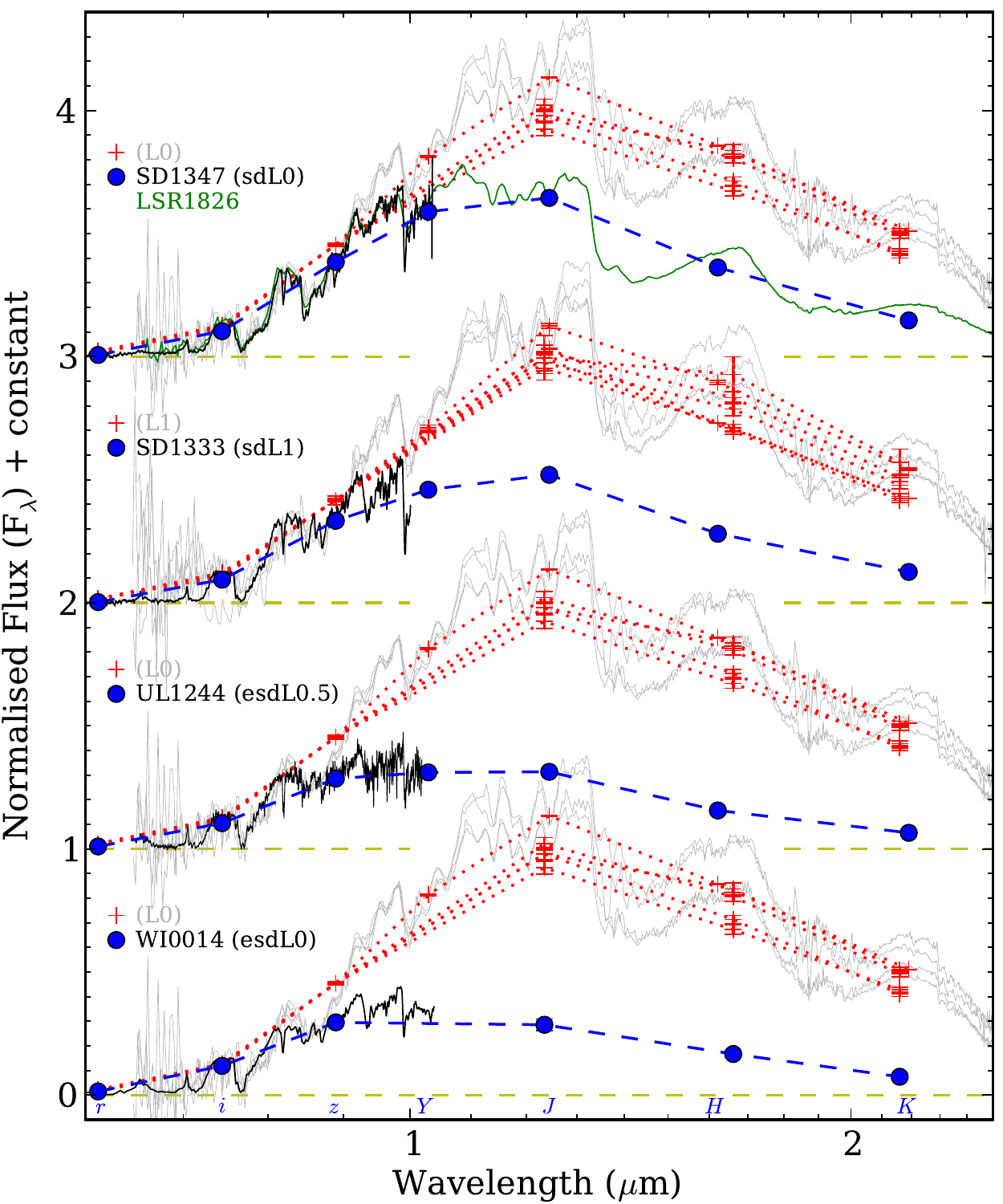}
\caption[]{SDSS--UKIDSS photometric flux points and optical spectra of four L subdwarfs (black) compared to Spex spectra of L0 dwarfs which are plotted as greyed out. The photometric flux points of each object are joined with dotted/dashed lines. The spectrum of LSR 1826+3014 \citep[LSR1826;][]{lep02} plotted in green is from \citet{bur04b}.  These L0 dwarfs are: 2MASS J12212770+0257198, 2MASSW J0228110+253738 \citep{bur08a}, 2MASSI J2107316--030733, and 2MASS J13313310+3407583 \citep{kir10}. These L1 dwarfs are: 2MASS J01340281+0508125 \citep{kir10}, 2MASSW J0208183+254253 \citep{bur08a}, SDSS J104842.84+011158.5 \citep{bur08a}, and 2MASS J20343769+0827009 \citep{burg10}.}
\label{3sdlsed}
\end{center}
\end{figure}

The 0.8 $\mu$m VO band is present in the spectra of late-type M and early-type L dwarfs \citep[e.g.,][]{boch07,zha09}. The 0.8 $\mu$m VO absorption band co-exists with the 2.3 $\mu$m CO absorption band in early-type sdL subdwarfs \citep[e.g. 2M1756;][]{kir10}, and is a strong indicator of metallicity. The VO band weakens as metallicity decreases, and eventually disappears. The top panel of Fig. \ref{mspec} shows optical spectra of early-type L subdwarfs with very different VO band strengths due to different metallicity. This effect is reproduced in the BT-Settl model \citep{alla14} spectra. Middle and bottom panels of Fig. \ref{mspec} show BT-Settl optical spectra  with $T_{\rm eff}$ of 2000 and 2400 K (corresponding to early and mid L types according to Section \ref{ssatmo}). Section \ref{ssmodel} shows that the best-fitting model parameters for SD1347 are $T_{\rm eff} = 2400 $K and $[Fe/H] = -0.5$  and  for SSS1013 are $T_{\rm eff} = 2700 $K and $[Fe/H] = -1.8$. The 0.77--0.81 $\mu$m region changes continuously with decreasing metallicity. We classify objects that have a weaker VO band (compared to L dwarfs) as sdL subdwarfs (e.g. 2M1756), and classify objects without a 0.8 $\mu$m VO absorption band as esdL. The 0.77--0.81 $\mu$m spectral profile of an early-type esdL should be well approximated by a straight slope. early-type L subdwarfs with significantly more flux in the 0.77--0.81 $\mu$m region should be classified as usdL to indicate an even more extreme effect, which is also contributed by a weakening of TiO absorption at 0.77 $\mu$m (as [Fe/H] changes from --1.5 to --2.5; see Fig. \ref{cahtiom}).

We classify early-type L subdwarfs by comparing their optical spectra to L dwarfs. 
Fig. \ref{2sdlop} shows the optical spectra of SD1333 and SD1347 compared to dwarf standards. The optical spectrum of SD1347 is very similar to L0, but there are slightly stronger CaH and TiO absorption bands, and the NIR photometric flux points are suppressed (see Fig. \ref{3sdlsed}). We thus classified SD1347 as sdL0. 
SD1333 was previously classified as sdL3 based on its low SNR SDSS spectrum in \citet{zha12}. Our new OSIRIS spectrum of SD1333 has a much better SNR ($\sim$150) and is very similar to the spectra of L0.5--L1 types. However, it  has also somewhat stronger CaH and TiO absorption bands, very weak 0.8 $\mu$m VO absorption, and largely suppressed NIR photometric flux points (Fig. \ref{3sdlsed}). We thus re-classify SD1333 as sdL1. \citet{kir16} also obtained a new optical spectrum of SD1333 and classified it as sdL0. Within the sdL subclass, SD1347 is relatively metal-rich and SD1333 is relative metal-poor, according to the strength of their 0.8 $\mu$m VO bands. 
Following the same strategy as for SD1347 and SD1333, we classify 2M0041, WI0014, and UL1244 as sdL0.5, esdL0, and esdL0.5, respectively (see Fig. \ref{3sdkn} and \ref{3sdlsed}).

Table \ref{tmetalc} presents a note summary of the spectral characteriztics of the L subdwarf metallicity subclasses that we have used to make our classifications.

\begin{table*}
 \centering
  \caption[]{Spectral characteriztics of the metallicity subclasses of L subdwarfs.}
  \begin{tabular}{c l l}
\hline\hline
Subclass    &  Spectral Characteriztics  & Examples   \\  
\hline
sdL & $H$ and  $K$ bands are more suppressed  than in L dwarfs  (normalizing in optical) & SD1416, UL0216  (Fig. \ref{l7yn}) \\
& CaH and TiO at around 0.7 $\mu$m are slightly deeper than in L dwarfs &2M1756 \citep{kir10}\\ 
& VO band at 0.8 $\mu$m in early-type sdL is weaker than in L dwarfs & 2M1756 \citep{kir10} \\
& 0.77--0.81 $\mu$m spectral profile of early-type esdL dips below a straight line & SD1333  (Fig. \ref{2sdlop}) \\
& FeH at 0.99 $\mu$m  in mid-late-type sdL is stronger than in L dwarfs  & SD1416  (Fig. \ref{l7yn})\\
& CO band at 2.3 $\mu$m  is weaker than in dL & 2M1756, SD1416  (Fig. \ref{l7yn})\\
& TiO at 0.85 $\mu$m stronger than for same spectral type L dwarfs & SD1347  (Fig. \ref{2sdlop}) \\  
\hline
esdL & $J, H$, and $K$ bands are strongly suppressed compared to L dwarfs (normalizing in optical). &  2M0616, UL1519  (Fig. \ref{l7yn})  \\
& CaH and TiO at around 0.7 $\mu$m are deeper than in L dwarfs & UL1244  (Fig. \ref{3sdkn}) \\
& VO band at 0.8 $\mu$m in early-type esdL disappears & WI0014, UL1244  (Fig. \ref{3sdkn})\\
& 0.77--0.81 $\mu$m spectral profile of early-type esdL well approximated by a straight slope & UL1244  (Fig. \ref{3sdkn}) \\  
& FeH at 0.99 $\mu$m in mid-late-type esdL is much stronger than in L dwarfs & 2M0616, 2M0532  (Fig. \ref{l7yn})\\
& CO band at 2.3 $\mu$m  disappears, $K$ band is almost flat & 2M0616, 2M0532  (Fig. \ref{l7yn})\\
& TiO at 0.85 $\mu$m weaker than same spectral type sdL & UL1244, 2M0616  (Fig. \ref{visspec}) \\  
\hline
usdL & $J, H$, and  $K$ bands are significantly suppressed compared to L dwarfs (normalizing in optical). & 2M1626  (Fig. \ref{l7yn})  \\
& CaH and TiO at around 0.7 $\mu$m are deeper than in dL &  SSS1013  (Fig. \ref{mspec}) \\
& VO band at 0.8 $\mu$m in early-type usdL disappears &  SSS1013  (Fig. \ref{mspec}) \\
& 0.77--0.81 $\mu$m spectral profile of early-type usdL appears well above a straight line & SSS1013  (Fig. \ref{mspec}) \\  
& FeH at 0.99 $\mu$m in mid--late-type usdL is much stronger than in L dwarfs & 2M1626  (Fig. \ref{l7yn}) \\
& CO band at 2.3 $\mu$m disappears, $K$ band is somewhat flat & 2M1626  (Fig. \ref{l7yn}) \\
& TiO at 0.85 $\mu$m weaker than same spectral type esdL & 2M1626 (Fig. \ref{visspec}) \\  
\hline
\end{tabular}
\label{tmetalc}
\end{table*}

\begin{table}
 \centering
  \caption[]{Known L subdwarfs.}
\label{tallsdl}
  \begin{tabular}{c c c  l }
\hline
    Name  &  SpT1$^{a}$ & Ref$^{b}$ & SpT2$^{c}$   \\
\hline
SSSPM J10130734$-$1356204 & sdM9.5 & 19,6 & usdL0 \\ 
SDSS J125637.13$-$022452.4 & sdL3.5 & 21,7 & usdL3  \\
ULAS J135058.86+081506.8 & sdL5 & 16 & usdL3 \\
2MASS J16262034+3925190 & sdL4 & 3 & usdL4  \\
WISEA J213409.15+713236.1 & sdM9 & 13 & usdL0.5  \\
\hline
WISEA J001450.17$-$083823.4 & sdL0 & 12,15 & esdL0  \\
WISEA J020201.25$-$313645.2 & sdL0 & 12 & esdL0.5  \\
WISEA J030601.66$-$033059.0 & sdL0 & 12,15 & esdL1  \\ 
ULAS J033350.84+001406.1 & sdL0 & 17 & esdL0  \\
WISEA J043535.82+211508.9 & sdL0 & 12,15 & esdL1  \\
2MASS J05325346+8246465 & sdL7 & 2 & esdL7  \\
2MASS J06164006$-$6407194 & sdL5 & 9 & esdL6  \\
ULAS J124425.90+102441.9 & sdL0.5 & 17 & esdL0.5 \\
SSSPM J144420.67$-$201922.2  & sdL0 & 18,13 & esdL1  \\
ULAS J151913.03$-$000030.0  & esdL4 & 1 &  esdL4  \\
2MASS J16403197+1231068 & sdM9/sdL & 4,10 & esdL0  \\
WISEA J204027.30+695924.1 & sdL0 & 12,15 & esdL0.5  \\
\hline
2MASS J00412179+3547133 & sdL? & 4 & sdL0.5  \\
WISEA J005757.65+201304.0 & sdL7 & 12,15 & ---  \\
WISEA J011639.05$-$165420.5 & d/sdM8.5 & 20 & sdL0 \\
WISEA J013012.66$-$104732.4 & d/sdM8.5 & 20 & sdL0 \\
ULAS J021642.97+004005.6    & sdL4 & 1 & sdL4   \\
2MASS J06453153$-$6646120 & sdL8 & 11 & ---  \\
WISEA J101329.72$-$724619.2 & sdL2? & 13 & ---  \\
2MASS J11582077+0435014 & sdL7 & 11 & --- \\
ULAS J124947.04+095019.8 & sdL1 & 1 & sdL1 \\
SDSS J133348.24+273508.8  & sdL1 & 1 & sdL1 \\ 
ULAS J133836.97$-$022910.7 &  sdL7 & 1  & sdL7 \\ 
SDSS J134749.74+333601.7   &sdL0 & 1 & sdL0 \\
WISEA J135501.90$-$825838.9 & sdL5? & 13 & ---  \\
SDSS J141624.08+134826.7 & d/sdL7 & 8,11 & sdL7  \\
2MASS J17561080+2815238 & sdL1 & 11 & ---  \\
LSR J182611.3+301419.1 & d/sdM8.5 & 14,4 & sdL0 \\
\hline
\end{tabular}
\begin{list}{}{}
\item[$^{a}$] Spectral types from the literature.
\item[$^{b}$] 1. This paper; 2. \citet{bur03}; 3. \citet{bur04}; 4. \citet{bur04b}; 5. \citet{bur06b}; 6. \citet{bur07}; 7. \citet{bur09}; 8. \citet{bur10}; 9. \citet{cus09}; 10. \citet{giz06};  11. \citet{kir10}; 12. \citet{kir14}; 13. \citet{kir16}; 14. \citet{lep02}; 15. \citet{luhm14};  16. \citet{lod10}; 17. \citet{lod12}; 18. \citet{sch04}; 19. \citet{sch04b}; 20. \citet{schn16}; 21. \citet{siv09}.
\item[$^{c}$] Spectral types adopted in this paper. Objects not examined in this paper have  no value here. 
\end{list}
\end{table}

\subsection{Spectral type of other known L subdwarfs}
\label{ssspt}

\begin{figure}
\begin{center}
   \includegraphics[width=\columnwidth]{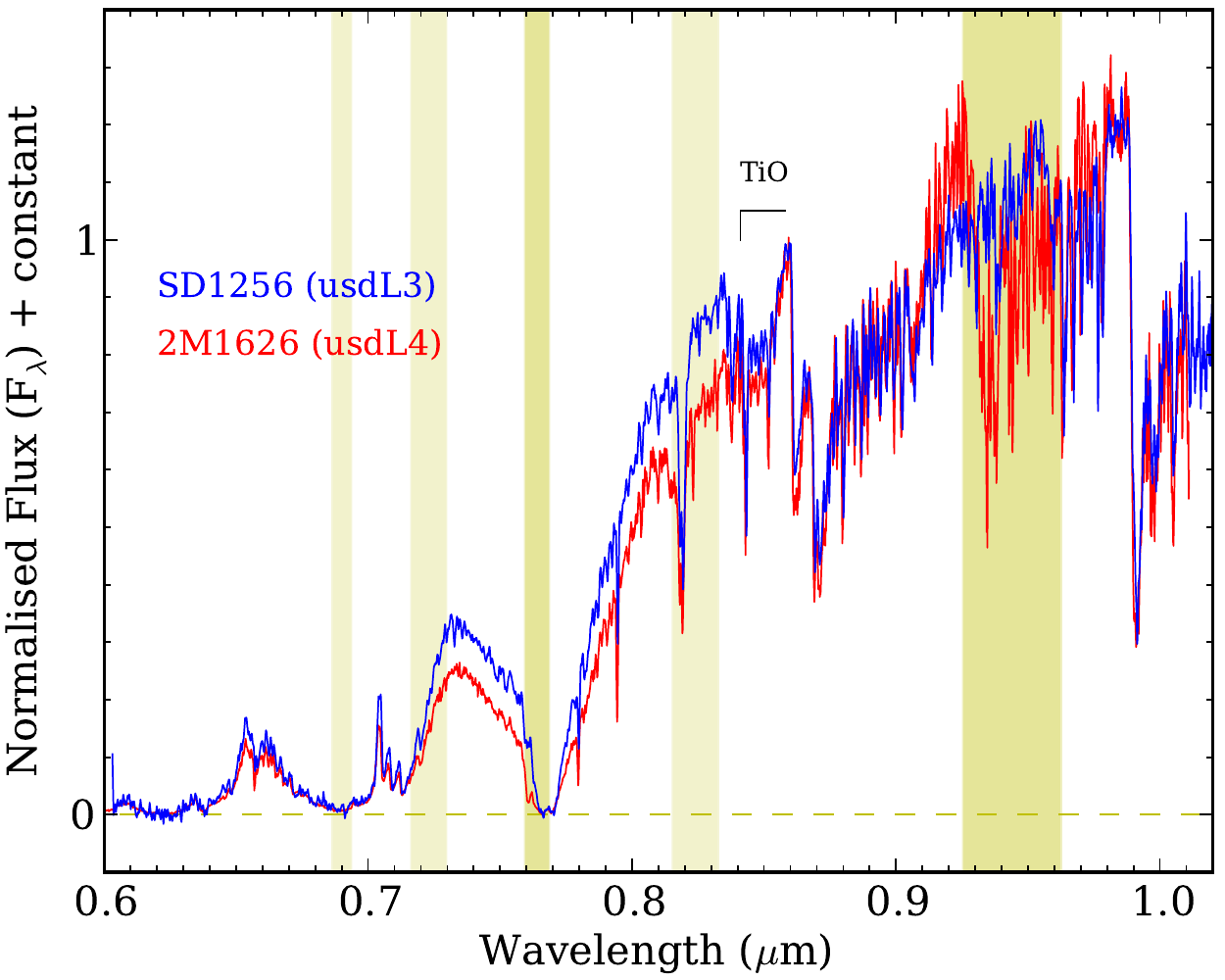}
\caption[]{Optical spectra of SD1256 and 2M1626. Spectra are normalised at 0.86 $\mu$m. 
Telluric absorptions are corrected.
}
\label{fsd1256}
\end{center}
\end{figure}

We have re-examined spectral types and subclasses of some known L subdwarfs: 2M0532 (esdL7), 2M0616 (esdL6), 2M1626 (usdL4), 2M0041 (sdL0.5), WI0014 (esdL0), and UL1244 (esdL0.5) in Section \ref{ssclass}. We also classified six new L subdwarfs: UL0216  (sdL4), UL1249 (sdL1), SD1333 (sdL1), UL1338 (sdL7), SD1347 (sdL0), and UL1519 (esdL4). Here we discuss the spectral types and spectral subclasses of other known blue L dwarfs and L subdwarfs based on the properties summarised in Table \ref{tmetalc}. 

Fig. \ref{mspec} shows that it is more and more difficult to assign spectral type to early-type L subdwarfs when $[Fe/H] < -1.5$ by direct comparison to optical spectra of L dwarfs. This is because TiO bands become very sensitive to metallicity and shape the spectra of early-type usdL subdwarfs in a way that is significantly different from L dwarfs. SDSS J125637.16$-$022452.2  \citep[SD1256;][]{siv09} was classified as sdL3.5 by \citet{bur09}. Its NIR spectrum has very similar properties as in 2M1626, i.e. flat in the $K$ band and 0.85 $\mu$m TiO absorption; thus, we classify it as an usdL subdwarf. Fig. \ref{fsd1256} shows that SD1256 has an optical spectrum that is significantly different from 2M1626, justifying that an SD1256 spectral type  is one subtype earlier than 2M1626. We Therefore, classify SD1256 as usdL3. 

ULAS J135058.86+081506.8  (UL1350) was classified as sdL5 by comparing its optical spectrum to those of 2M1626 and 2M0616 \citep[see Figure 2. of][]{lod10}. If one only examines the spectrum at 0.7--0.9 $\mu$m in Figure 2. of \citet{lod10}, UL1350 is much more similar to SD1256 or 2M1626 than to 2M0616. The spectrum of UL1350 beyond 0.9 $\mu$m may not be reliable due to low SNR and/or poor second-order flux calibration. UL1350 is not plotted in Fig. \ref{ijk} because it will overlap with SD1256 as they have identical $i-J$ and $J-K$ colours. We Therefore, classify UL1350 as usdL3. 

The 0.8 $\mu$m VO absorption is absent in spectra of early-type esdL subdwarfs like SD1244 and WI0014. Other known objects have this feature including: 
SSSPM J144420.67$-$201922.2 \citep[SSS1444; fig. 2. of][]{sch04}, 2MASS J16403197+1231068 \citep[2M1640; fig. 9. of][]{bur07}, ULAS J033350.84+001406.1 \citep[UL0333; fig. 4. of][]{lod12}, and WISEA J020201.25$-$313645.2 (WI0101), WISEA J030601.66$-$033059.0 (WI0306), WISEA J043535.82+211508.9 (WI0435), and WISEA J204027.30+695924.1 (WI2040) in fig. 25 of \citet{kir14}. Thus we proposed to classify these objects as esdL.

\begin{figure}
\begin{center}
   \includegraphics[width=\columnwidth]{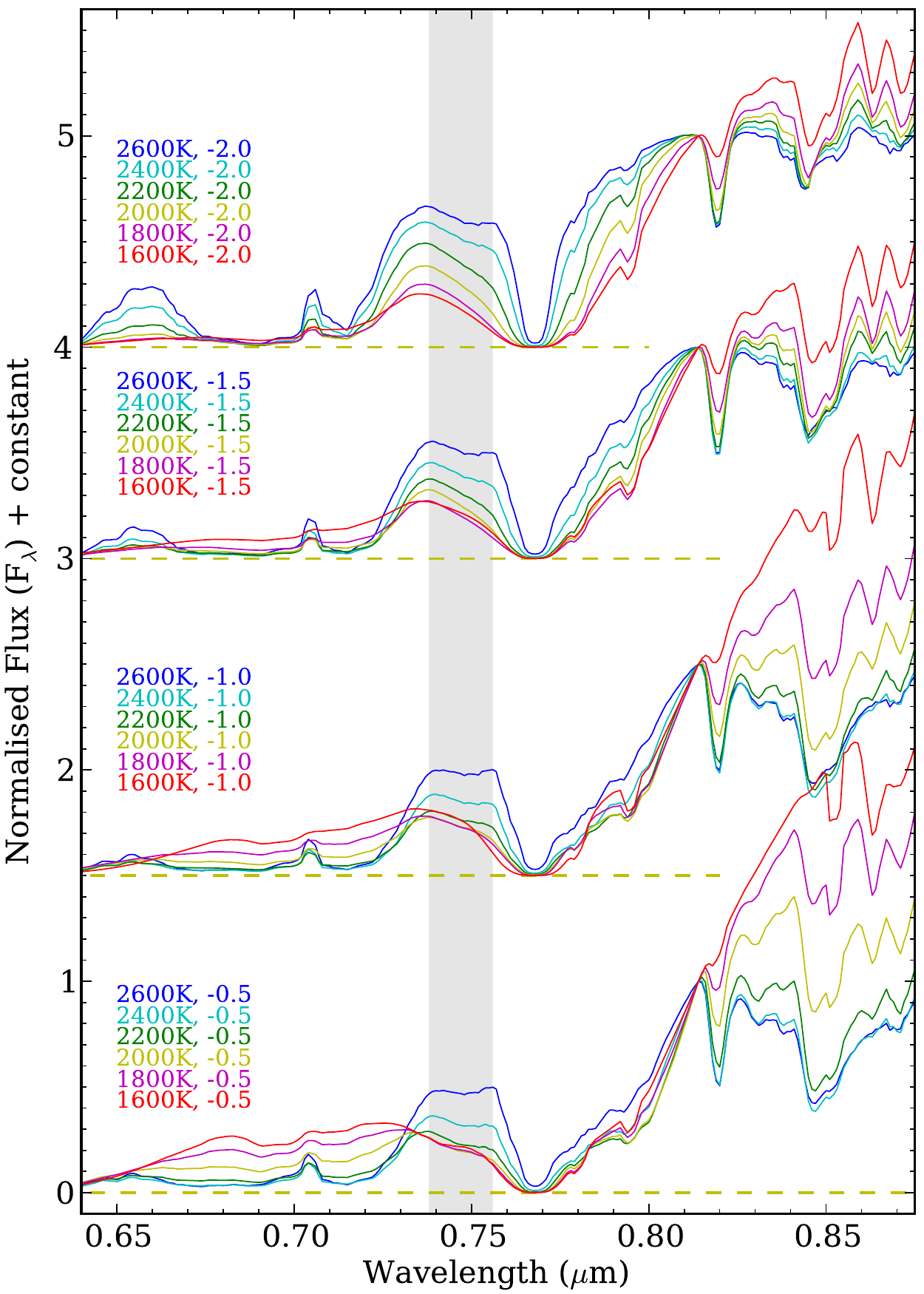}
\caption[]{Comparison of BT-Settl optical spectra with different $T_{\rm eff}$. All spectra have  log $g$ =  5.5.  $T_{\rm eff}$ and [Fe/H] are labelled above each set of spectra. Spectra are normalised at 0.815 $\mu$m. }
\label{mspecteff}
\end{center}
\end{figure}

By comparing the optical spectra of known late-type M and early-type L subdwarfs, \citet{kir14} discovered that there is a plateau at 0.738--0.757 $\mu$m that can be used to assign spectral types of L subdwarfs. The slope at the top of this plateau slowly changes from slightly redward to flat through the sdM9--sdL0.5 sequence, then becomes blueward for sdL1. This phenomenon is reproduced by the BT-Settl models \citep{alla14}. Fig. \ref{mspecteff} shows that this spectral slope (light yellow shaded region) changes continuously across the $T_{\rm eff}$ = 2600--1600 K region.

WI0014, WI0202, WI2040, WI0306, and WI0435 discovered by \citet{kir14} have plateaus with flat or slightly blueward slopes and were classified as sdL0. We classify these objects as esdL subdwarfs as we discussed earlier in this section. If we re-examine the spectra in fig. 25 of \citet{kir14}, we find that WI0202 and WI0240 actually have 0.738--0.757 $\mu$m plateaus as flat as UL1244, thus suggesting esdL0.5. Although WI0306 and WI0435 have different metallicity subclass to 2M1756, they all have blueward plateaus, and we thus classify WI0306 and WI0435 as esdL1. Fig. \ref{3sdkn} shows that WI0014 has an almost flat plateau but has a dip around 0.756 $\mu$m, and we thus classify it as esdL0.

2M1640 has similar spectrum as UL0333, which suggests it is also an esdL0 \citep[see fig. 9. of][]{bur07}.  SSS1444 has similar spectrum to WI0306 and WI0435, which suggests they should have an esdL1 classification \citep[see fig. 2. of][]{sch04}. 

SSS1013 (Fig. \ref{mspec}) was classified as esdM9.5 by \citet{bur07}.  The 0.738--0.757 $\mu$m plateau of this object appears fairly flat but with a dip at 0.76 $\mu$m. The 0.77-0.81 $\mu$m profile of SSS1013 is significantly above a straight line slope (due to weakening of 0.77 $\mu$m TiO), which indicates an usdL subclass. Therefore, we classify SSS1013 as usdL0. WISEA J213409.15+713236.1 (WI2134) was classified as sdM9 \citep[fig. 63.][]{kir16}. Its 0.738-0.757 $\mu$m plateau appears somewhat flat, suggesting a later type than sdM9. The 0.77-0.81 $\mu$m profile of WI2134 is significantly above a straight line slope (similar to SSS1013) indicating an usdL subclass, and we thus classify WI2134 as usdL0.5. 

LSR1826 was classified as d/sdM8.5 from its NIR spectrum by \citet{bur04b}. Fig. \ref{2sdlop} shows that LSR1826 has the same optical spectrum as SD1347, and we thus classify it as sdL0. WISEA J011639.05$-$165420.5 and WISEA J013012.66$-$104732.4 in fig. 12. of \citet{schn16} compare well with LSR1826, and we thus classify them as sdL0. 

Table \ref{tallsdl} shows a list of currently known L subdwarfs with updated spectral types. 16 are sdL, 12 are esdL and 5 are usdL.

\subsection{Enhancement and suppression for the different L dwarf subclasses}
\label{ssspecp}

\begin{figure}
\begin{center}
   \includegraphics[width=\columnwidth]{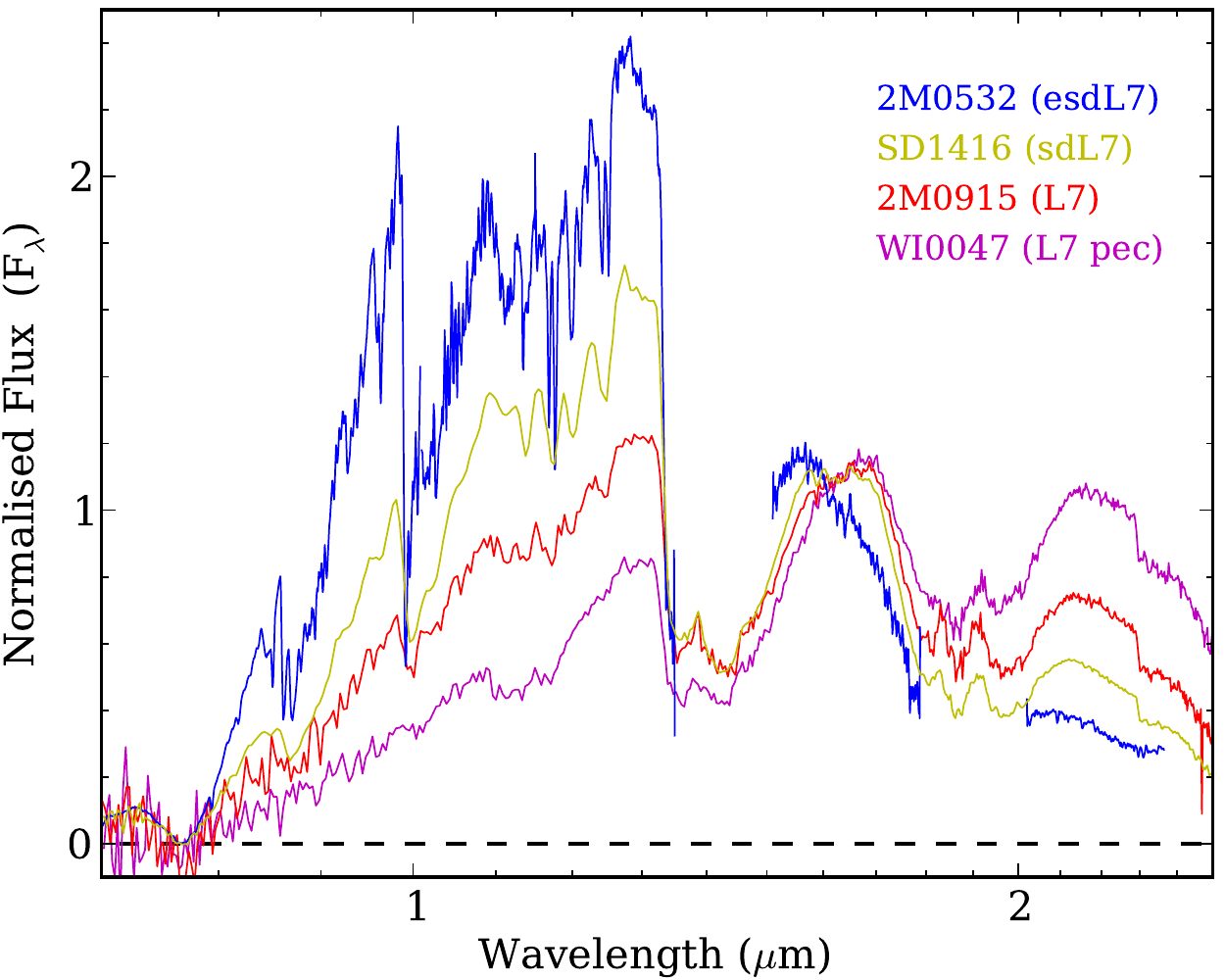}
\caption[]{Spectra of L7 dwarfs/subdwarfs normalised in the $H$ band at 1.6 $\mu$m. WISEP J004701.06+680352.1 (WI0047) is from \citet{giz12}.}
\label{l7hn}
\end{center}
\end{figure}

\begin{figure}
\begin{center}
   \includegraphics[width=\columnwidth]{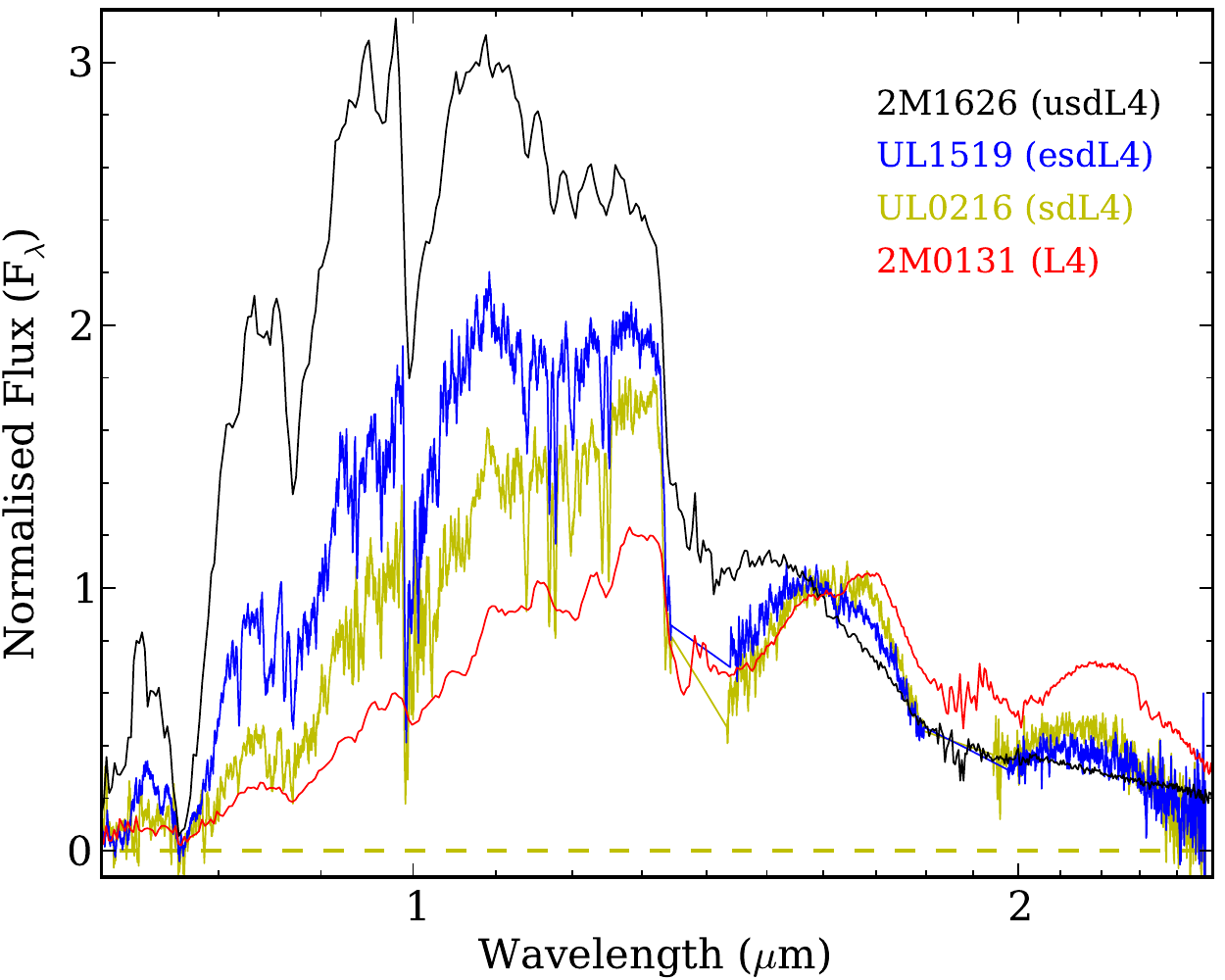}
\caption[]{Spectra of L4 dwarfs/subdwarfs normalised in the $H$ band at 1.6 $\mu$m. }
\label{l4hn}
\end{center}
\end{figure}

To consider relative enhancement/suppression for the different L dwarf subclasses, we plot Fig. \ref{l7hn} and Fig. \ref{l4hn}. Fig. \ref{l7hn} shows spectra for a confined range of $\sim$L7 spectral type spanning a range of spectral peculiarity and subclass. Objects in this spectral type range should all be BDs. To give an indication of relative flux levels the spectra are normalised at 1.6 $\mu$m in Fig. \ref{l7hn}. Since L dwarfs/subdwarfs have similar $M_H$ magnitudes (see Figure 3 of \citealt{zha13}), this plot indicates relative brightness levels in an absolute sense. The full sequence runs through; L7 pec, L7, sdL7, and esdL7. At the two extremes WI0047 is a  young ($\sim$ 0.1 Gyr) and low-mass BD \citep[$\sim$ 19 M$_{\rm Jup}$;][]{giz15}, while 2M0532 is an old ($\goa$ 10 Gyr) and massive BD  \citep[$\sim$ 80 M$_{\rm Jup}$;][]{bur08b}. The redistribution of flux from longer to shorter wavelength leads to higher absolute flux level for subdwarfs at 0.8--1.4 $\mu$m.  It is also clear that in addition to differences in metallicity, a large spread in mass (and log $g$) and age is also evident for any particular spectral type. Similar to Fig. \ref{l7hn}, Fig. \ref{l4hn} shows spectra of L4, sdL4, esdL4, and usdL4 normalised at 1.6 $\mu$m. It is obvious that an usdL4 subdwarf would have a much warmer $T_{\rm eff}$ than an L4 dwarf according to their SED.

\begin{table*}
 \centering
  \caption[]{Coefficients of third-order polynomial fits of absolute magnitude ($M_{\rm abs}$) as a function of spectral types ($SpT$) for M0--L7 subdwarfs in Fig. \ref{fmjh}. The fits are defined as  $M_{\rm abs} = c_0 + c_1 \times {\rm SpT} + c_2 \times {\rm SpT}^{2} + c_3 \times {\rm SpT}^{3}$. SpT = 0 for M0 and SpT = 10 for L0.  The root mean square (rms) of polynomial fits are listed in the last column. }
 \label{tmjh}
  \begin{tabular}{l c c c c c}
\hline
    $M_{\rm abs}$  & $c_0$ & $c_1$ & $c_2$ & $c_3$ & rms (mag) \\
\hline
$M_{\rm J}$ (MKO)       & 8.64788 & $3.17384\times10^{-1}$ & $-1.76459\times10^{-2}$ & $8.53625\times10^{-4}$ & 0.40  \\ 
$M_{\rm H}$ (MKO)      & 8.19731 & $2.71013\times10^{-1}$ & $-4.54248\times10^{-3}$ & $2.90020\times10^{-4}$ & 0.40 \\
$M_{\rm J}$ (2MASS)   & 8.68342 & $3.16187\times10^{-1}$ & $-1.75984\times10^{-2}$ & $8.48172\times10^{-4}$ & 0.40 \\
$M_{\rm H}$ (2MASS)  & 8.18494 & $2.81607\times10^{-1}$ & $-7.53663\times10^{-3}$ & $4.32261\times10^{-4}$ & 0.41 \\
\hline
\end{tabular}
\end{table*}

\begin{table}
 \centering
  \caption[]{Astrometry, distance and radial velocities of our six new L subdwarfs.}
 \label{tsdpmplx}
  \begin{tabular}{c c c c c}
\hline
    Name  & $\mu_{\rm RA}$ & $\mu_{\rm Dec}$  & Distance$^{a}$ & RV \\
      & (mas yr$^{-1}$)  & (mas yr$^{-1}$)  & (pc) & (km s$^{-1}$) \\
\hline
 UL0216 & $-$61$\pm$8 & $-$98$\pm$8  & 103$^{+21}_{-17}$ & $-$90$\pm$14 \\
 UL1249 & $-$243$\pm$11 & $-$212$\pm$6 & 119$^{+24}_{-20}$  & $-$176$\pm$32  \\
 SD1333 & 103$\pm$6 & $-$604$\pm$6  & 112$^{+23}_{-19}$ & 48$\pm$30 \\
 UL1338 & $-$48$\pm$4 & $-$261$\pm$8 & 60$^{+12}_{-10}$  & $-$136$\pm$38 \\
 SD1347 & 70$\pm$12 & $-$16$\pm$9 &  88$^{+18}_{-15}$ & $-$83$\pm$7 \\
 UL1519 & $-$22$\pm$10 & $-$421$\pm$10 & 108$^{+22}_{-18}$  & 80$\pm$14     \\
\hline
\end{tabular}
\begin{list}{}{}
\item[$^{a}$] Spectroscopic distances based on the relationship between spectral type and $H$ band absolute magnitudes  (Fig. \ref{fmjh}). 
\end{list}
\end{table}

\subsection{Kinematics of L subdwarfs}
\label{suvw}

Dwarf stars orbit the Galactic Centre in a similar direction as part of the Galactic thin disc, while cool subdwarfs may be part of the (more dispersed) thick disc or could be on more extended orbits within the Galactic halo. Thus cool subdwarfs will generally have more dispersed $U, V$ and $W$ space velocities compared to dwarfs \citep[$U$ is positive in the direction of the Galactic anti centre, $V$ is positive in the direction of galactic rotation, and $W$ is positive in the direction of the North Galactic Pole; ][]{joh87}. The $U, V, W$ space velocity components are thus indicators for membership of the different Galactic populations.

\begin{figure}
\begin{center}
   \includegraphics[angle=0,width=\columnwidth]{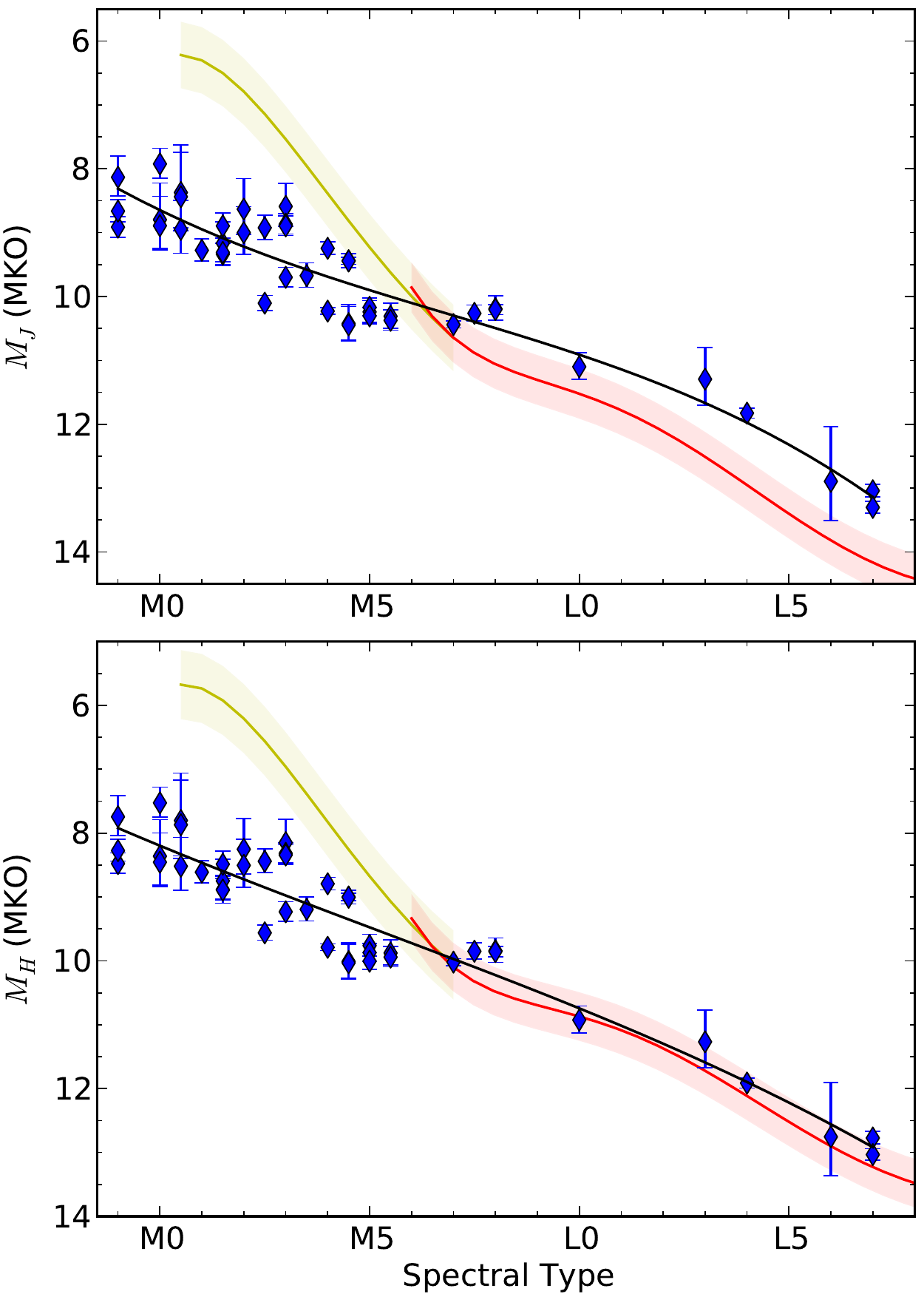}
\caption[]{Relationships of spectral types and $J$- and $H$-band absolute magnitudes of M and L subdwarfs updated from \citet{zha13}, which are plotted as black lines in both panels.  The relationships for M0.5--M7 dwarfs (yellow lines) from \citet{zha13} and M6--L dwarfs (red lines) from \citet{dup12} are plotted for comparison. Shaded areas show their fitting rms.} 
\label{fmjh}
\end{center}
\end{figure}

We calculated $U, V, W$ space velocities for L subdwarfs based on their distances, radial velocities (RV) and proper motions following \citet{clar10}. Proper motions were calculated based on SDSS and UKIDSS astrometry. To measure the spectroscopic distances of our objects we updated the spectral type versus absolute magnitude relationships in \citet{zha13}. Fig. \ref{fmjh} shows the spectral type and absolute magnitude relationships for $M_{\rm J}$ and $M_{\rm H}$ in MKO photometry. Table \ref{tmjh} shows the coefficients of polynomial fits to these relationships in both MKO and 2MASS photometric systems.  These relationships are fitted with M and L subdwarfs of esd and usd subclasses.  From Fig. \ref{fmjh} we can see that the spectral type and $M_{\rm H}$ relationships of dwarfs and subdwarfs are very similar between M7 and L7. This means M7--L7 subdwarfs of different metallicity subclasses have similar $M_{\rm H}$ if they have same subtypes. Therefore, we estimated distances of our objects with the spectral type and $M_{\rm H}$ relationship which minimised the uncertainty due to subclass classification. 
The radial velocities of UL0216, UL1519, UL1249 and UL1338 were measured using their K {\scriptsize I} lines in the $J$ band, while radial velocities of SD1333 and SD1347 were calculated from redshifts in the SDSS data base (based on cross-correlated SDSS spectroscopy). Table \ref{tsdpmplx} presents the distance constraints and astrometric measurements for our six new L subdwarfs. 

\begin{figure*}
\begin{center}
   \includegraphics[angle=0,width=\textwidth]{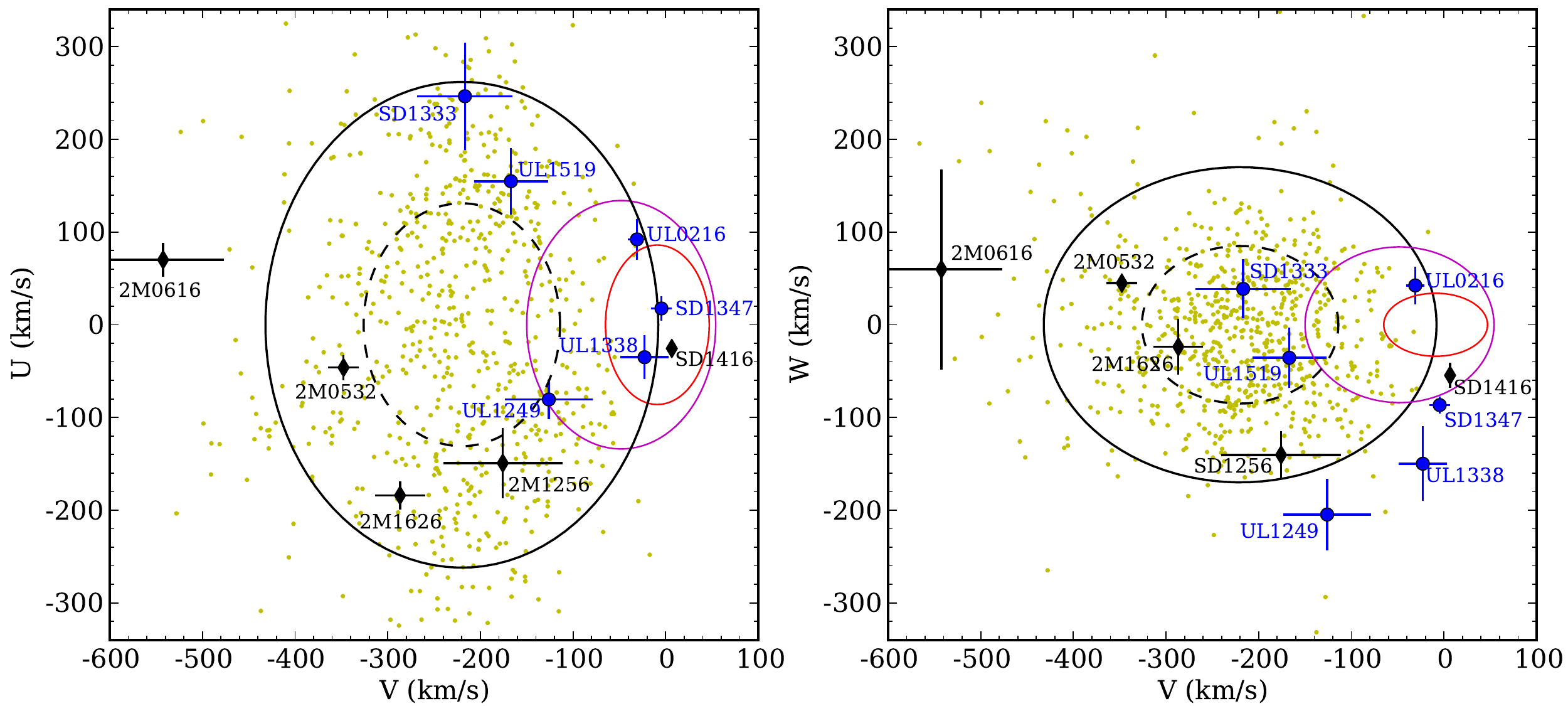}
\caption[]{$U, V, W$ space velocities of 11 L subdwarfs. Blue filled circles represent the six new L subdwarfs reported here. Black diamonds represent five known L subdwarfs with parallax distances.  The red, magenta and black solid lines are 2$\sigma$ velocity dispersions of the Galactic thin disc, thick disc and halo, respectively \citep{red06}. The black dashed line is the 1$\sigma$ velocity dispersion of the halo. Yellow dots are esdM and usdM subdwarfs from  \citet{zha13}.} 
\label{uvwsdl}
\end{center}
\end{figure*}

Fig. \ref{uvwsdl} shows the $U, V, W$ velocities of 11 subdwarfs including 5 known L subdwarfs with parallax distances, and our 6 new L subdwarfs. The 1$\sigma$ and 2$\sigma$ velocity dispersions of the thin disc, thick disc and halo \citep{red06}, and esdM and usdM subdwarfs \citep{zha13} are also plotted for comparison. While expected scatter in velocity precludes direct kinematic association of individual objects, we can usefully consider the overall kinematic distribution in Fig. \ref{uvwsdl}. It can be seen that none of the L subdwarfs (previously known and new) lie within the 2$\sigma$ thin disc velocity dispersions in both plots. Four out of five of the previously known L subdwarfs lie beyond the 2$\sigma$ thick disc velocity dispersion, whereas approximately 50 per cent of the new L subdwarfs lie in this region. This is consistent with the L subdwarfs being members of the thick disc or halo populations. It is also indicative (though these are low number statistics) of the new sample having a somewhat higher fraction of thick-disc members (compared to halo members).

\section{Atmospheric properties}
\label{smodel}

\subsection{Model comparison}
\label{ssmodel}

Optical--NIR spectra of L subdwarfs are affected by $T_{\rm eff}$, metallicity, and log $g$ in a complicated way. The NIR spectra are mainly affected by $T_{\rm eff}$ and metallicity, and less so by log $g$. But the optical spectra are most sensitive to $T_{\rm eff}$, with a lower level of metallicity and log $g$ sensitivity. Thus, taken together the optical--NIR model comparisons combine to provide an improved ability to yield $T_{\rm eff}$ and metallicity constraints of L subdwarfs. Although the broadness of the K {\scriptsize I} wings is gravity sensitive, this is not detrimental to L subdwarf classification since they are all old and have small variation in surface gravity.

Atmospheric models can reproduce the overall observed SED of UCSDs, and can closely reproduce a variety of optical and NIR spectral features. For model fitting we made use of the BT-Settl model grids\footnote{\url{https://phoenix.ens-lyon.fr/Grids/BT-Settl/}}. The BT-Settl model grids for 2700 K $\lid T_{\rm eff} \lid$ 3000 K are from \citet{alla11}, cover $-2.5 \lid$ [Fe/H] $\lid -0.5$ and 5.0 $\lid$ log $g$ $\lid$ 5.5, with intervals of 100 K for $T_{\rm eff}$ and 0.5 dex for both [Fe/H] and log $g$. The BT-Settl model grids for 1400 K $\lid T_{\rm eff} \lid$ 2600 K are from \citet{alla14}, cover  $-2.5 \lid$ [Fe/H] $\lid -0.5$ and 5.0 $\lid$ log $g$ $\lid$ 5.75, with intervals of 100 K for $T_{\rm eff}$, 0.5 dex for  [Fe/H] and 0.25 dex for log $g$ (surface gravity). We also used linear interpolation between some models where this was able to yield an improved fit.

We took a non-standard approach to fitting these models. Non uniform levels of fit quality (across different wavelength features), and the availability of model grid coverage, make routine reduced-Chi-squared ($\chi^2$) fitting problematic. We Therefore, adopted a hybrid method (combining visual fits with uncertainty estimates informed by reduced $\chi^2$ calculations). We identified best-fitting BT-Settl model spectra through visual comparison with our observed spectra, noting (see below) any outstanding issues with our chosen best fits. Our output fit results include BT-Settl model parameters where a favourable comparison was found. To assess the uncertainties associated with these fits we selected a representative test-sample from amongst our subdwarfs, and measured the reduced $\chi^2$ values for their best-fitting models. We then determined reduced $\chi^2$ values for models with parameters close to the best-fitting (where model grid availability is allowed), and used linear interpolation to estimate parameter uncertainties representative of $\pm 1\sigma$ (i.e. a reduced $\chi^2$ increase of 1.0). The results were reasonably uniform across our test sample, and indicate uncertainties of $\sim$120 K in $T_{\rm eff}$, $\sim$0.2 dex in [Fe/H], and $\sim$0.2 dex in log $g$.

To provide an additional test for the models and check the reliability of our results, we performed our fits not only for the 6 new subdwarfs (Table \ref{tsdlm}) and 3 known L subdwarfs that we observed, but also for another 13 known late-type M and L subdwarfs for which optical and NIR spectra were available. Table \ref{tmodel} shows the resulting best fit atmospheric parameters for all 22 UCSDs.

\begin{figure*}
\begin{center}
  \includegraphics[width=\textwidth]{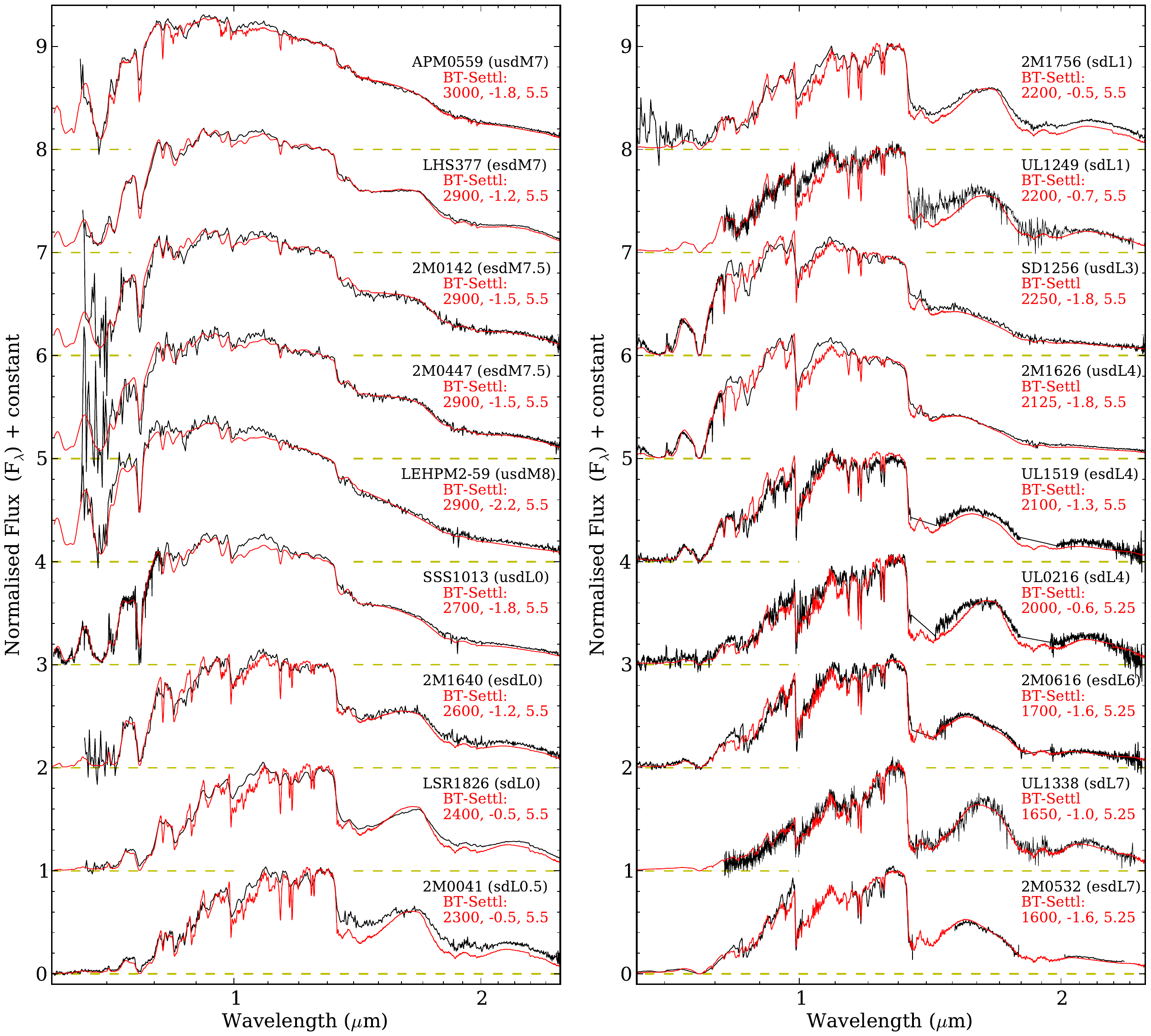}
\caption[]{Optical--NIR spectra of 18 late-type M and L subdwarfs compared to the best-fitting BT-Settl models. $T_{\rm eff}$, [M/H] and log $g$ of the models are indicated. Spectra are normalised at 1.3 $\mu$m. Model spectra have resolving power of  1000 for 1600 K $\lid T_{\rm eff} \lid 2000$ K, 500 for 2100 K $\lid T_{\rm eff} \lid 2600$ K and 200 for 2700 K $\lid T_{\rm eff} \lid 3000$ K at 1 $\mu$m. 
Spectra of APM0559 and LEHPM 2-59 are from \citet{bur06b}; LHS377 and SSSPM1013 are from \citet{bur04}; 2M0041, 2M0142, and 2M1640 are from \citet{bur04b}; and 2M0447 is from \citet{kir10}. The optical spectrum (0.65--0.82 $\mu$m) of SSS1013 is from \citet{bur07}. The spectrum of 2M0616 and  the optical spectrum (0.6--0.92 $\mu$m) of 2M0041 are from this paper. Spectra of 2M0532 and 2M1756 are from \citet{bur03} and \citep{kir10}, respectively. Optical spectra (before 0.82 $\mu$m) of SD1256 and 2M1626 are from \citet{bur09} and \citet{bur07}, respectively.}
\label{sdmlmodel}
\end{center}
\end{figure*}

\begin{figure}
\begin{center}
  \includegraphics[width=\columnwidth]{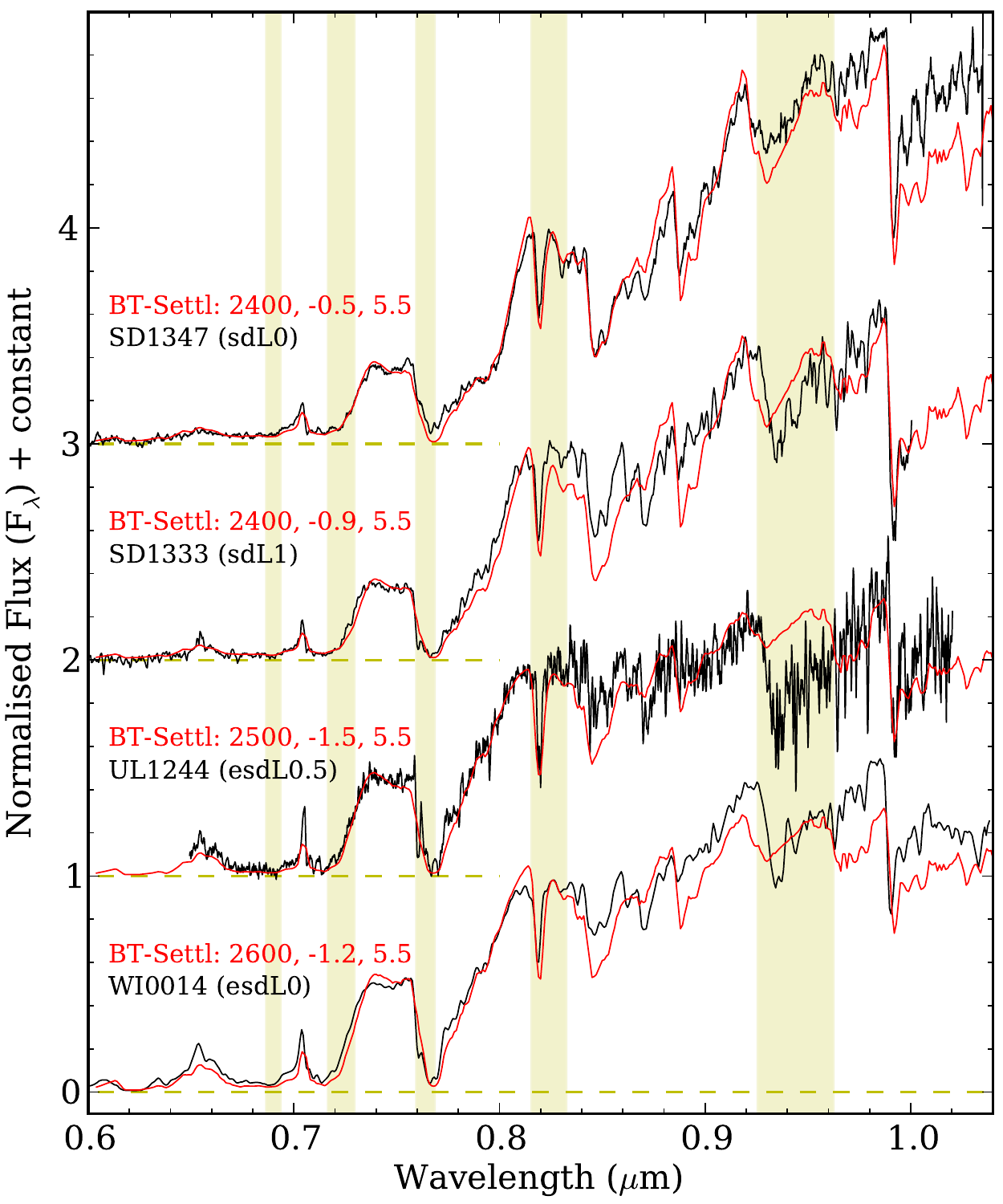}
\caption[]{Optical spectra of four subdwarfs compared to the best fitting BT-Settl models. $T_{\rm eff}$, [Fe/H], and log $g$ of the models are indicated. 
Spectra are normalised at 0.815 $\mu$m. Model spectra have a resolving power of 500 at 1 $\mu$m. }
\label{4sdlmodel}
\end{center}
\end{figure}

\begin{table*}
 \centering
  \caption[]{Atmospheric properties of 22 UCSDs (Figs \ref{sdmlmodel} and \ref{4sdlmodel}) derived from BT-Settl models.  SpT1 is the spectral type in the literatures and SpT2 is the spectral type adopted in this paper. These six L subdwarfs have no value on SpT1 and reference are new. The metallicity parameter in the PHOENIX models is defined as iron abundance, thus [M/H] indicated in models is equivalent to [Fe/H]. } 
 \label{tmodel}
  \begin{tabular}{l c c c c l l l}
\hline
    Name & Short name  & $T_{\rm eff} (K)$ & [Fe/H] & log $g$ & SpT1 & Reference & SpT2   \\
\hline
APMPM 0559-2903 & APM0559 & 3000 & --1.8  & 5.50 & esdM7 & \citet{bur06b} & usdM7 \\
LHS 377 & --- & 2900 & --1.2  & 5.50 & sdM7 & \citet{bur04} & esdM7 \\
2MASS J01423153+0523285 & 2M0142 & 2900 & --1.5  & 5.50 & sdM8.5 & \citet{bur04b} & esdM7.5 \\
2MASS J04470652-1946392 & 2M0447 & 2900 & --1.5 & 5.50 & sdM7.5 & \citet{kir10} & esdM7.5 \\
LEHPM 2-59 & --- & 2900 & --2.2 & 5.50  & esdM8 & \citet{bur06b} & usdM8 \\
SSSPM 1013-1356 & SSS1013 & 2700 & --1.8  & 5.50  & esdM9.5 & \citet{bur04} & usdL0 \\
2MASS J16403197+1231068 & 2M1640 & 2600 & --1.2 & 5.50  & sdM9/sdL & \citet{giz06} & esdL0 \\
WISEA J001450.17-083823.4 & WI0014 & 2600 & --1.2 & 5.50 & sdL0 & \citet{kir14} & esdL0\\
ULAS J124425.75+102439.3 & UL1244 & 2500 & --1.5  & 5.50 & sdL0.5 & \citet{lod12} & esdL0.5 \\
LSR 1826+3014 & LSR1826 & 2400 & --0.5  & 5.50 & d/sdM8.5 & \citet{bur04b} & sdL0 \\
 2MASS J00412179+3547133 & 2M0041 & 2300 & --0.5 & 5.50 & sdL & \citet{bur04b} & sdL0.5 \\
SDSS J134749.74+333601.7 &  SD1347 & 2400 & --0.5 & 5.50  & --- & --- & sdL0 \\
SDSS J133348.24+273508.8 & SD1333 & 2400 & --0.9   & 5.50 &  --- & --- & sdL1 \\
 ULAS J124947.04+095019.8  & UL1249 & 2200 & --0.7 & 5.50 & --- & --- & sdL1   \\ 
2MASS J17561080+2815238  & 2M1756 & 2200 & --0.5 & 5.50 & sdL1 & \citet{kir10}  & sdL1 \\
 SDSS J125637.16$-$022452.2 & SD1256  & 2250 & --1.8 & 5.50 & sdL3.5 & \citet{bur09} & usdL3  \\
 2MASS J16262034+3925190 & 2M1626 & 2125 & --1.8  & 5.50 & sdL4 & \citet{bur04} & usdL4 \\
 ULAS J151913.03$-$000030.0 & UL1519 & 2100 & --1.3 & 5.50  & --- & --- & esdL4 \\
 ULAS J021642.97+004005.6 & UL0216 & 2000 & --0.6  & 5.25 & --- & --- & sdL4 \\
  2MASS J06164006$-$6407194 & 2M0616 & 1700 & --1.6 & 5.25 & sdL5 & \citet{cus09} & esdL6 \\
ULAS J133836.97$-$022910.7 & UL1338 & 1650 & --1.0  & 5.25 & --- & --- & sdL7 \\
 2MASS J05325346+8246465 & 2M0532 & 1600 & --1.6  & 5.25 & sdL7 & \citet{bur03} & esdL7 \\
 \hline
\end{tabular}
\end{table*}

Fig. \ref{sdmlmodel} shows optical+NIR spectra of late-type M and L subdwarfs compared to BT-Settl models. Fig. \ref{4sdlmodel} shows the optical spectra of four L subdwarfs (SD1347, SD1333, UL1244 and WI0014) for which no NIR spectral coverage was available. Overall 22 late-type M and L subdwarfs were fitted well by BT-Settl models.

From Fig. \ref{sdmlmodel} we can see that the BT-Settl model fits of very metal-poor UCSDs (e.g. $[Fe/H] < -1.5$) are better than for objects with higher metallicity. This is possibly because more metal-poor atmospheres are simpler and easier to model. UL0216 is fitted well by the BT-Settl model spectrum with $T_{\rm eff}$=1600 K, $[Fe/H] = -0.6$, and log $g$ = 5.25. However, the model over estimates the water absorption band around 1.5 $\mu$m. The BT-Settl model fit to UL1519 is better in the optical, than in the NIR. UL1338 is fitted well by the model, but the BT-Settl models are not reliable at $T_{\rm eff} < 1800$ K and $[Fe/H] > -1.0$ when we consider the $J-K$ colours derived from the model spectra (Fig. \ref{ijk}). The model may not represent the true atmospheric parameters of UL1338. 

Fig. \ref{4sdlmodel} shows that optical spectra alone can provide reasonable results when fitting the properties of such early-type L subdwarfs. Further evidence for this comes from SD1347 (optical-only fit) and LSR1826 (optical + NIR fit, Fig. \ref{sdmlmodel}), which are fit well by the same model. 
The BT-Settl model was very effective at reproducing the observed spectrum of UL1244.

\subsection{Spectral type and $T_{\rm eff}$ relationships}
\label{ssatmo}

\begin{figure}
\begin{center}
   \includegraphics[width=\columnwidth]{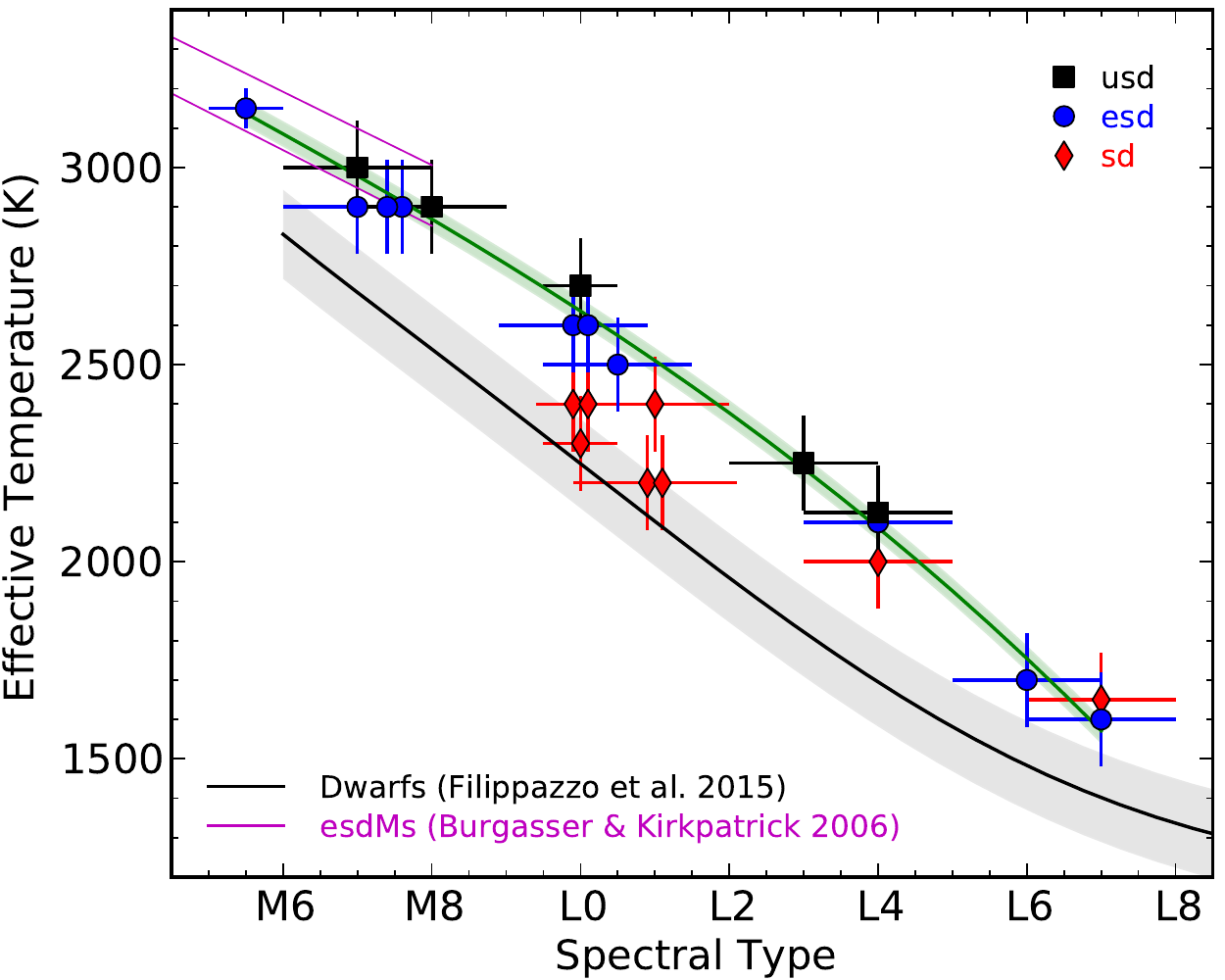}
\caption[]{Spectral types and $T_{\rm eff}$ of late-type M and L subdwarfs. The black line shows the spectral type and $T_{\rm eff}$ correlation from \citet{fili15} with an rms of 113 K (shaded area). Two purple solid lines are spectral type and $T_{\rm eff}$ correlations for esdMs based on optical (upper) and NIR (lower) spectra from \citet{bur06b}. The green solid line is our polynomial fit to the esd and usd subdwarfs [equation (7)] with an rms of 32.5 K (shaded area). Spectral subtypes are offset by $\pm$0.1 for clarity when two objects share the same spectral type and $T_{\rm eff}$.}
\label{sdmlt}
\end{center}
\end{figure}

The $T_{\rm eff}$ is typically the most important factor in shaping the spectra of VLMS and BD. Mid to late-type M subdwarfs are found to have higher $T_{\rm eff}$ than M dwarfs of the same type \citep{bur06b,raj14}. Fig. \ref{sdmlt} shows the relationship between spectral types and $T_{\rm eff}$ of late-type M and L subdwarfs provided in Table \ref{tmodel}. The errors on $T_{\rm eff}$ shown in Fig. \ref{sdmlt} are about 120 K. The $T_{\rm eff}$ values for these subdwarfs are about 100--400 K higher than dwarfs with the same spectral types. The $T_{\rm eff}$ of early-type sdL subdwarfs are about 100-200 K higher than early-type L dwarfs. Fig. \ref{sdmlt} also shows that a subdwarf can have similar $T_{\rm eff}$ to a dwarf classified 2--3 subtypes earlier. For instance, objects with spectral types of L0.5, sdL1 and usdL3 would have similar $T_{\rm eff}$. 
We have determined a polynomial fit to the spectral type (SpT) and $T_{\rm eff}$ of objects with esdM5.5--esdL7 and usdM7--usdL4 types, which follows: 
\begin{equation}
T_{\rm eff} = 3706 - 107.8 \times {\rm SpT} + 1.686 \times {\rm SpT}^2 - 0.1606 \times {\rm SpT}^3		
\end{equation}
with an rms of 32.5 K. In this equation SpT = 10 for esdL0/usdL0, and SpT = 17 for esdL7/usdL7 (etc). All sdLs were excluded in the fit simply because most of these examples are confined to a small range (sdL0--1) in the spectral subtype.  

Our $T_{\rm eff}$ estimates for late-type M subdwarfs are consistent with the results from \citet{bur06b} where they made NIR spectral fits to the subsolar metallicity models  NextGen \citep{hau99,all01} and \citet{ack01}. The $T_{\rm eff}$ of the four esdM5--esdM8 subdwarfs in \citet{bur06b} were estimated based on optical spectra and are about 150 K higher than those based on NIR spectra. The $T_{\rm eff}$ of late-type M subdwarfs estimated from high-resolution optical spectra and BT-Settl models in \citet{raj14} is also 150--200 K higher than our results. Thus, there is a discrepancy between the $T_{\rm eff}$ difference (between late-type M subdwarfs and dwarfs) reported by \citet{raj14} and that found in our analysis (400--500 K and 200--400 K, respectively). Also, Fig. 7 of \citet{bur06b} presents a $T_{\rm eff}$ difference (between the sequences) of 400-600 K, based on NIR spectral fits. The difference with our result is mainly due to the M dwarf $T_{\rm eff}$ scale that we used \citep{fili15}, which is warmer than that used by \citet{bur06b}. The older spectral type $T_{\rm eff}$ relation for M dwarfs underwent some improvement by \citet{fili15}, who used a larger sample and newer models. This work is also consistent with a sample of M dwarfs from \citet{mann15} for which $T_{\rm eff}$ estimation were relatively independent of models.

\section{Discussions}
\label{sdisc}

\subsection{Metallicity ranges of the subclasses of M and L subdwarfs}
\label{smetal}

Metallicity plays an important role in shaping the spectra of VLMS and BD, causing shifts in the spectral types and temperature scale. L subdwarfs are a natural extension of M subdwarfs into lower mass and $T_{\rm eff}$ regimes. 
M subdwarfs are brighter and more numerous than L subdwarfs, and relatively well characterized; thus, they provide a useful comparison and possible reference for the characterization of L subdwarfs.

To determine the metallicity subclasses of M dwarfs and subdwarfs LRS07 used the metallicity index $\zeta_{\rm TiO/CaH}$, and defined four metallicity subclasses: ultra subdwarf (usdM; $\zeta_{\rm TiO/CaH} < 0.2$), extreme subdwarf  (esdM; $0.2 < \zeta_{\rm TiO/CaH}  < 0.5$), subdwarf (sdM; $0.5 < \zeta_{\rm TiO/CaH} < 0.825$) and dwarf (dM; $\zeta_{\rm TiO/CaH} > 0.825$). The metallicity distributions of these four subclasses became clear when  metallicity measurements were made based on optical high-resolution spectra \citep[e.g.][]{wool09}. This allowed a relationship (albeit with a scatter) to be established between $\zeta_{\rm TiO/CaH}$ and iron abundance, which was recently refined by \citet{pavl15} who combined data from \citet{wool06} and \citet{wool09} to give
\begin{equation}
{\rm [Fe/H]} = 2.00 \times \zeta_{\rm TiO/CaH} - 1.89
\end{equation}
with an rms of 0.26. 
However, equation (8) is valid only for early-type M subdwarfs, because all the objects in the Woolf sample are M0--M3 subdwarfs.

We calculated approximate metallicity ranges for the four LRS07 subclasses of M0--M3 subdwarfs using the $\zeta_{\rm TiO/CaH}$ ranges from LRS07 and equation (8) (these are presented in the left-hand side of Table \ref{tmetal}). As we discussed in Section \ref{ssmsd}, the metallicity consistency of $\zeta_{\rm TiO/CaH}$ is tested only for early-type M subdwarfs. The $\zeta_{\rm TiO/CaH}$ index is not a consistent indicator of metallicity across all M subtypes and L types.

\begin{figure}
\begin{center}
   \includegraphics[width=\columnwidth]{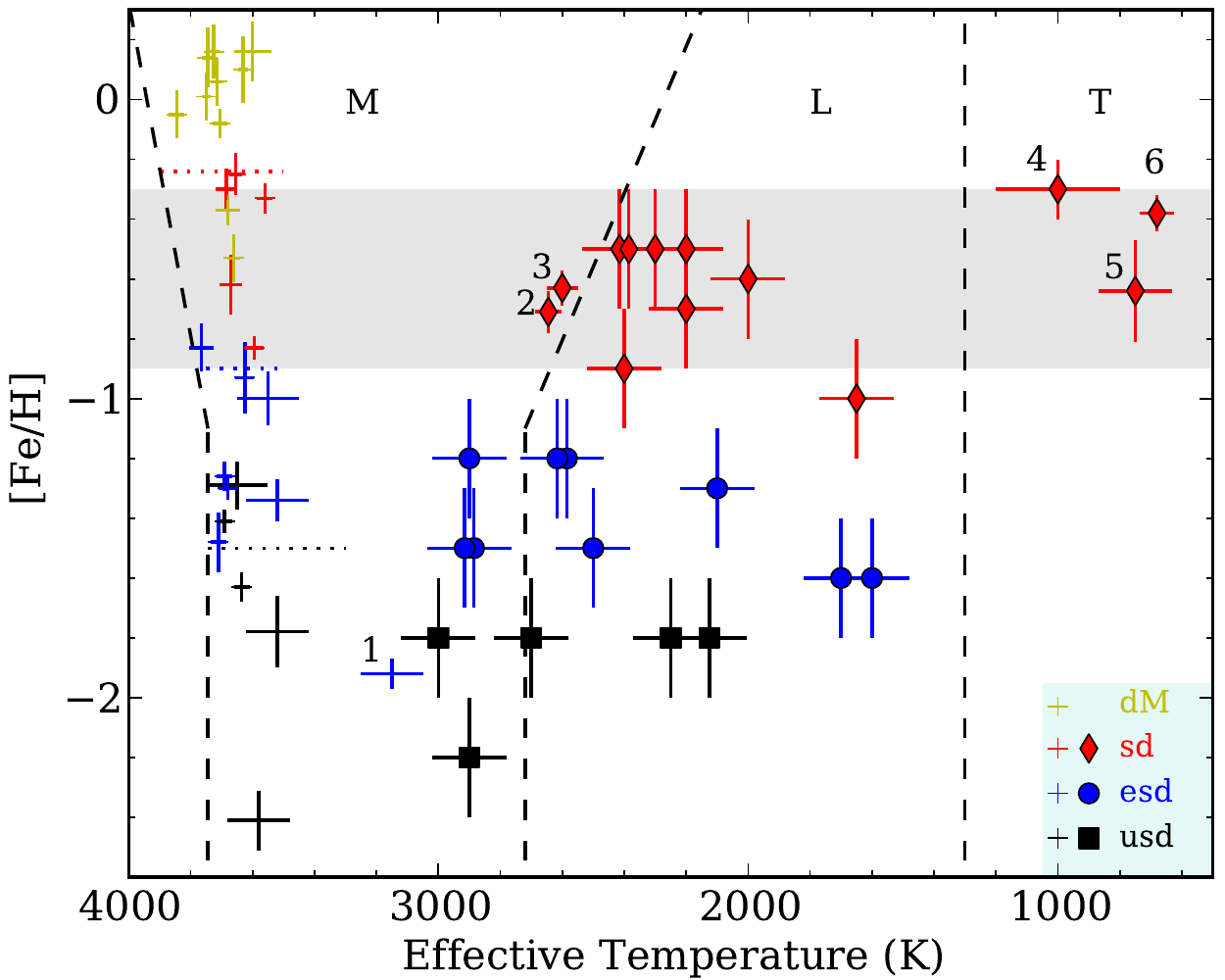}
\caption[]{[Fe/H] and $T_{\rm eff}$ of M, L, and T subdwarfs. Black dashed lines indicate the boundaries between K, M, L, and T types. Horizontal red, blue, and black dotted lines indicate [Fe/H] boundaries (Table \ref{tmetal}) between early-type dM, sdM, esdM, and usdM derived from fig. 9 of \citet{pavl15}. Objects with $T_{\rm eff} >$ 3500 K (yellow, red, blue and black crosses are for dM, sdM, esdM, and usdM, respectively) are from \citet{wool06} and \citet{wool09}. Objects labelled with numbers `1--6' have metallicity measurements inferred from their primary stars. `1'  is G224-58 B  \citep{pavl15};  `2' is HD 114762 B  \citep{bow09}; `3' is GJ 660.1 B \citep{agan16}; `4' is Hip 73786 B \citep[T6p;][]{murr11};  `5' is WISE 2005+5424 \citep[sdT8;][]{mac13}; and `6' is BD+01$^{\circ}$ 2920 B \citep[T8p;][]{pin12}.  The remaining $T_{\rm eff} <$ 3000 K objects are provided in Table \ref{tmodel}. The shaded area indicates the rough [Fe/H] range for the thick disc population, with the thin disc population above and the halo population below. The $T_{\rm eff}$ of some objects has been offset by $\pm$15 K for clarity, if they share the same $T_{\rm eff}$ and [Fe/H] as another object.}
\label{mostmp}
\end{center}
\end{figure}

Fig. \ref{mostmp} explores how metallicity subclass distributions map on to the metallicity-$T_{\rm eff}$ plane for M, L and T types. Three black dashed lines indicate the boundaries between K, M, L and T dwarfs/subdwarfs which are derived from spectral type--$T_{\rm eff}$ relationships of late-type M and L dwarfs \citep{fili15} and subdwarfs [equation (7)] augmented with data from \citet{mann15}. 
Different symbol shapes/colours indicate different spectral subclasses (see  caption of Fig. \ref{mostmp}). 
These late M and L subdwarf subclasses are modified from the literature in Section \ref{ssclass}. We note that there are no L subdwarf benchmark companions currently known, and although there are additional known T subdwarfs in the literature, none have metallicity constraints as robust as the objects shown in the plot. 

The approximate metallicity ranges of the subclasses of M0--M3 defined by LRS07 are shown as dotted lines in the left side of the plot. It can be seen that these metallicity ranges reasonably bracket the four LRS07  metallicity subclasses (d, sd, esd, and usd), though there is some scatter that leads to each LRS07  subclass spreading into adjacent metallicity ranges (this will be discussed further later in this section). We also establish the approximate metallicity ranges for the subclasses of L subdwarfs (or more generally the $T_{\rm eff} \lid$ 3000 K population). The metallicity range for these UCSDs is [Fe/H] $> -0.3$ and is $-1.0 <$ [Fe/H] $\lid -0.3$ for the sd subclass. These are very similar to the metallicity ranges of the LRS07 dM0-3 and sdM0-3 subclasses. At lower metallicity (for $T_{\rm eff} \lid$ 3000 K), the metallicity range is $-1.7 <$ [Fe/H] $\lid -1.0$  for the esd subclass and is $[Fe/H] \lid -1.7$ for the usd subclass. These cover slightly different metallicity ranges than the (M0--M3) LRS07 esdM and usdM subclasses. 

By comparison, the kinematic halo population of F, G, and K stars have $[Fe/H] \loa -0.9$ and a metallicity distribution function peaks at $[Fe/H] \approx -1.7$ \citep{lair88,spa10,an13}, well matched to the two lowest metallicity ranges for both classification schemes. And thin disc stars generally have $[Fe/H] > -0.3$ \citep[e.g. from APOGEE; ][]{hay15}, well matched to the highest metallicity range for both schemes. 

Although the metallicity ranges for the two subclass schemes appear reasonably consistent, there is some evidence that they may not be consistent in the late M regime. The metallicity ranges of the LRS07 subclasses were estimated using  M0--M3 subdwarfs, and we note three later dwarfs in the LRS07 esdM subclass that have metallicity well below the approximate range expected from M0-M3 dwarfs. G224-58 B (esdM5.5 according to LRS07) has a significantly lower metallicity than earlier esdM dwarfs, and APM0559 and LEHPM 2-59 have similarly low metallicity and are classified as esdM by LRS07 and usdM in this paper. Changing metallicity ranges within a metallicity subclass is not ideal, and attempts to mitigate against this were made by LRS07 through the use of wide binary systems (whose components should have common metallicity) to help define subclass divisions. However, the lack of subdwarf binaries with early and late M components could have led to metallicity gradients across the LRS07 subtypes. Any such gradients appear to be largely absent from the $T_{\rm eff} <$ 3200 K subclasses scheme. Clearly more binary systems like SDSS J210105.37--065633.0 AB \citep[esdM1.5+esdM5.5;][]{zha13,pavl15} would be very useful if the metallicity subclasses of early-late M subdwarfs are to be refined. Table \ref{tmetal} summarises both subclass schemes, and indicates approximate links between subclasses, metallicity and kinematic populations.

 \begin{table*}
 \centering
  \caption[]{Metallicities ranges of subclasses of early-type M and L dwarfs/subwarfs. }
  \begin{tabular}{r  c c c c r  c}
\hline\hline
$^{a}$Subclass  & [Fe/H] & | & Kinematics & | & Subclass &   [Fe/H] \\  
\hline
dM0-3 &   $ > -$0.24 & | & Thin disc &| & dL & $ > -$0.3   \\
sdM0-3  & $ (-0.9, -0.24]$ & |& Thick disc &| & sdL & $ (-1.0, -0.3]$   \\ 
esdM0-3 & $ (-1.5, -0.9]$  & |& Halo &| & esdL & $ (-1.7, -1.0]$    \\
 usdM0-3 &   $\lid -$1.5   & |& Halo &| & usdL & $\lid -$1.7    \\
\hline
\end{tabular}
\begin{list}{}{}
\item[$^{a}$] Metallicity subclasses of M dwarfs/subdwarfs are based on the classification scheme of LRS07. 
\end{list}
\label{tmetal}
\end{table*}

\subsection{Absolute magnitudes of L and T subdwarfs}

\begin{figure}
\begin{center}
   \includegraphics[width=\columnwidth]{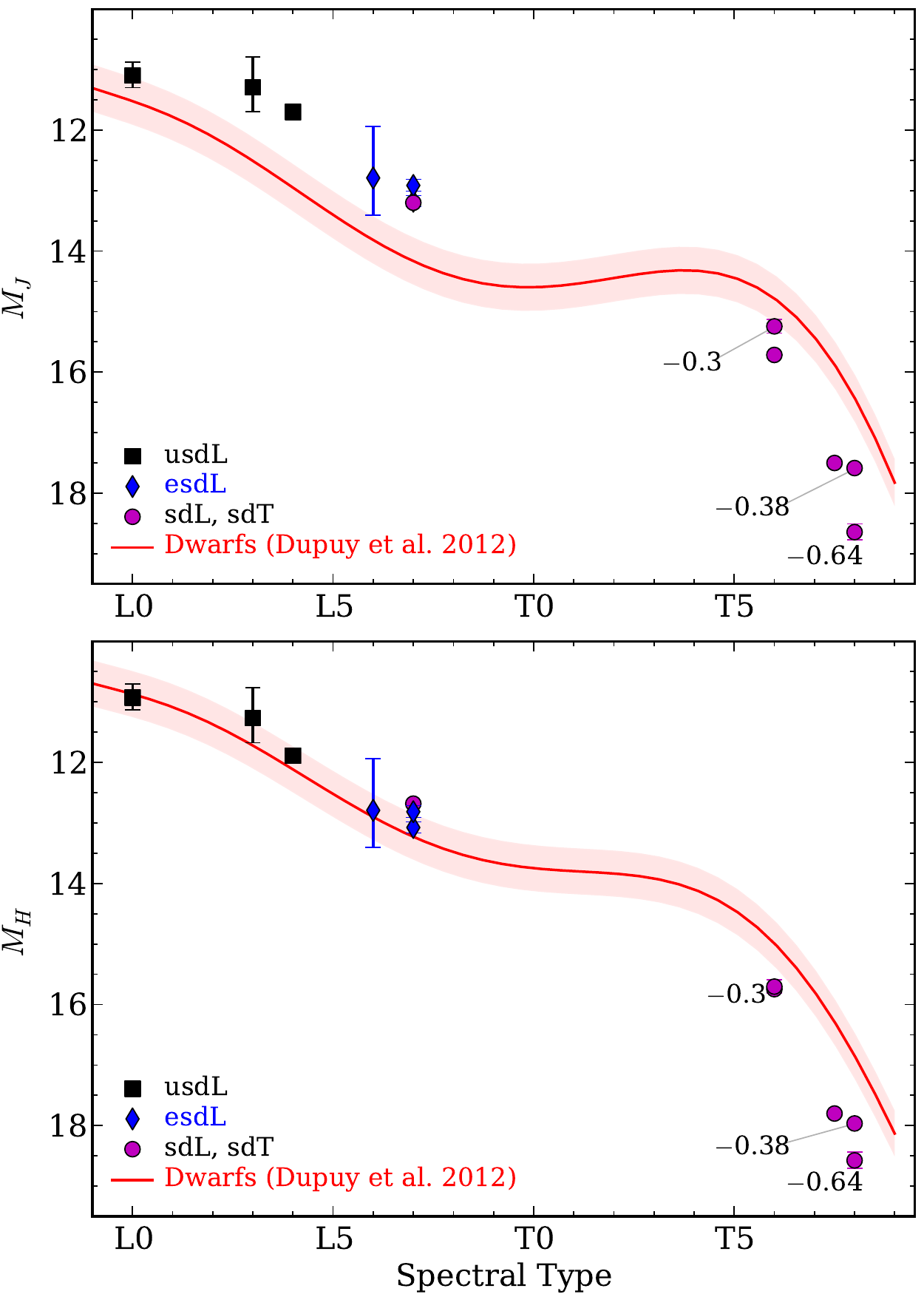}
\caption[]{The relationship between spectral type and $J$- and $H$-band absolute magnitudes (MKO) for L and T subdwarfs. The red solid line is for M--L--T dwarfs \citep{dup12}. The shaded area shows the fitting rms.  Three numbers to the left of three sdT companions indicate that [Fe/H] was inferred from their bright primary stars \citep{cena07,roja12,pin12}. 
Note  that sdL7 and  sdT7.5 are components of a wide binary SD1416 AB. Error bars for some objects are similar to or smaller than the plotting symbols. }
\label{fsptmj}
\end{center}
\end{figure}

In Fig. \ref{fsptmj} we plot $M_J$ and $M_H$ absolute magnitude against spectral type relationships for L and T dwarfs and subdwarfs. The dwarf sequence (red line) comes from \citet{dup12}. These six L subdwarfs with parallax distances are: 2M0532 \citep{bur08b,sch09},  2M0616 \citep{fah12},  SSS1013, 2M1256, and 2M1626 \citep{sch09}, and SD1416 A \citep{dup12}. To extend the subdwarf sequence into the T dwarf regime, we collected T subdwarfs with direct or indirect parallax measurements from the literature. They are either single objects with parallax distances or companions to bright stars which have parallax distances. The parallax of 2MASS J09373487+2931409 \citep[T6p;][]{bur02} was measured by \citet{sch09}. The parallax of SD1416 B \citep[T7.5p;][]{bur10} was from SD1416 A \citep{dup12}. The parallaxes of Hip 73786 B \citep[T6p;][]{murr11}, BD+01$^{\circ}$ 2920 B \citep[T8p;][]{pin12}, and WISE 2005+5424 \citep[sdT8;][]{mac13} are measured from their primary stars \citep{van07}. 

It is interesting to compare the dwarf and subdwarf sequences. M0-M5 dwarfs are brighter in the $J$ band than subdwarfs of the same spectral type, while M7-L7 dwarfs are fainter in the $J$ band (see Fig. \ref{fmjh}). Fig. \ref{fsptmj} shows that T dwarfs are brighter in $J$ and $H$ band than sdT subdwarfs of the same spectral type. A larger sample of L and T subdwarfs with parallax distances would allow us to have a better idea of how and why they are different from dwarfs. 

The sdT subdwarfs have $M_J$ and $M_H$ that are fainter by 1--2 mag when compared to T dwarfs with the same NIR spectral type. Distances of isolated late-type T subdwarfs will be over estimated by 2$\pm$0.5 times, if they are based on relationships between spectral type and $J$ or $H$ absolute magnitude \citep[e.g.][]{dup12,fah12}. \citet{pin14} also noted that the distance constraints (estimated from T dwarf absolute magnitude versus spectral type relations) for two highly $K$ band suppressed fast moving T subdwarfs are much greater when using NIR bands than for mid-infrared bands.

\section{Summary}
\label{ssumm}

In this paper we presented the discovery of six L subdwarfs from SDSS and UKIDSS (UL0216, UL1249, SD1333, UL1338, SD1347, and UL1519). We also presented new optical spectra of three previously known L subdwarfs (WI0014, 2M0041, and UL1244). We discussed the spectral properties of the known L subdwarfs, performed some re-classification of some known objects, and determined spectral type and subclass for our new L subdwarfs. 

We compared the nine measured objects with BT-Settl model spectra, and estimated their $T_{\rm eff}$ and metallicity. We also estimated atmospheric properties of another 13 known late-type M and L subdwarfs for which red optical and NIR spectra are available. BT-Settl models were successful in reproducing the overall optical--NIR spectral profile of M and L subdwarfs, particularly at $[Fe/H] \lid -1.0$. However, the BT-Settl models could not reproduce, in detail, some optical spectroscopic features of L subdwarfs. Our model fit results show that esdL and usdL subdwarfs have temperatures about 200--300 K higher than L dwarfs with the same spectral type, and have similar $T_{\rm eff}$ to L dwarfs that are about 2--3 subtypes earlier. 

We also found that the approximate metallicity ranges of the $T_{\rm eff} \lid$ 3000 K subclasses (including the L subdwarfs and some sdT dwarfs) are: $[Fe/H] \lid -1.7$ for usd, $-1.7 < [Fe/H] \lid -1.0$ for esd, and  $-1.0 < [Fe/H] \lid -0.3$ for sd. 
The metallicity ranges of the subclasses of cooler ($T_{\rm eff} <$ 3000 K) M and L subdwarfs are reasonably consistent with early-type M subdwarfs. However, there is some evidence for a metallicity gradient across the LRS07 subclasses. Binary systems containing both early- and late-type M subdwarfs could be an important tool if the $T_{\rm eff} >$ 3000 K M classification scheme is to be refined.

In the NIR, L subdwarfs are more luminous than L dwarfs with the same spectral type, while late-type sdT subdwarfs are less luminous than T dwarfs with the same spectral type. The $J$ band absolute magnitudes of five known late-type sdT subdwarfs are 1--2 mag fainter than T dwarfs with the same spectral type. Spectroscopic distances of known sdT subdwarfs would be over estimated by 2$\pm$0.5 times if based on spectral type and NIR absolute magnitude relationships for T dwarfs.

\section*{Acknowledgments}
This paper includes data gathered with the 6.5 meter Magellan Telescopes located at Las Campanas Observatory, Chile. Based on observations made with the Gran Telescopio Canarias (GTC), installed in the Spanish Observatorio del Roque de los Muchachos of the Instituto de Astrof{\'i}sica de Canarias, in the island of La Palma. 
This work is based in part on data obtained as part of the UKIRT Infrared Deep Sky Survey. 
The UKIDSS project is defined in \citet{law07}. UKIDSS uses the UKIRT Wide Field Camera \citep[WFCAM;][]{casa07}. The photometric system is described in \citet{hew06}, and the calibration is described in \citet{hodg09}. The pipeline processing and science archive are described in \citet{irwi04} and \citet{hamb08}.
Funding for the SDSS and SDSS-II has been provided by the Alfred P. Sloan Foundation, the Participating Institutions, the National Science Foundation, the U.S. Department of Energy, the National Aeronautics and Space Administration, the Japanese Monbukagakusho, the Max Planck Society, and the Higher Education Funding Council for England. The SDSS Web Site is http://www.sdss.org/. Funding for SDSS-III has been provided by the Alfred P. Sloan Foundation, the Participating Institutions, the National Science Foundation, and the U.S. Department of Energy Office of Science. The SDSS-III web site is http://www.sdss3.org/. This publication makes use of data products from the Two Micron All Sky Survey. 

This research has made use of the VizieR catalogue access tool, CDS, Strasbourg, France. Research has benefited from the M, L, and T dwarf compendium housed at DwarfArchives.org and maintained by Chris Gelino, Davy Kirkpatrick, and Adam Burgasser. This research has benefited from the SpeX Prism Spectral Libraries, maintained by Adam Burgasser at http://www.browndwarfs.org/spexprism. This publication makes use of VOSA, developed under the Spanish Virtual Observatory project supported from the Spanish MICINN through grant AyA2008-02156. 

ZHZ is supported by the IAC fellowship. ZZ and NL are partially funded by the Spanish Ministry of Economy and Competitiveness (MINECO) under the grants AYA2015-69350-C3-2-P and AYA2010-19136. ZHZ was partly supported by the Royal Astronomical Society to attend international conferences. ZHZ, BB, HRAJ, FM, RLS and JG have received support from the Marie Curie 7th European Community Framework Programme grant n.247593 Interpretation and Parametrization of Extremely Red COOL dwarfs (IPERCOOL) International Research Staff Exchange Scheme. DJP, BB, HRAJ, FM, ACD and JG are supported by STFC grant ST/M001008/1. FA and AJB were supported by the Jesus Ferra Foundation for their visit to the Instituto de Astrof{\'i}sica de Canarias. MCGO acknowledges the financial support of a JAE-Doc CSIC fellowship co-funded with the European Social Fund under the  programme {\em`Junta para la Ampliaci\'on de Estudios'} and the support of the Spanish Ministry of Economy and Competitiveness through the project AYA2014-54348-C3-2-R. FA received funding from the French `Programme National de Physique Stellaire' (PNPS) and the `Programme National de Plan\'etologie' of CNRS (INSU). DH is supported by Sonderforschungsbereich SFB 881 `The Milky Way System' (subproject A4) of the German Research Foundation (DFG).  The computations of atmosphere models were performed at the {\sl P\^ole Scientifique de Mod\'elisation Num\'erique} (PSMN) at the {\sl \'Ecole Normale Sup\'erieure} (ENS) in Lyon, and at the {\sl Gesellschaft f{\"u}r Wissenschaftliche Datenverarbeitung G{\"o}ttingen} in collaboration with the Institut f{\"u}r Astrophysik G{\"o}ttingen. 
The authors thank the referee, J. Davy Kirkpatrick for the useful and constructive comments.

\end{document}